\documentclass[review, dvipsnames]{elsarticle}

\usepackage{defaults}

\definecolor{positivestance}{rgb}{0.00000,0.44700,0.74100}
\definecolor{negativestance}{rgb}{0.85000,0.32500,0.09800}

\journal{Chaos, Solitons \& Fractals}

\begin{document}

    \title{An evidence-accumulating drift-diffusion model of competing information spread on networks}

    \author[1]{Julien Corsin} \ead{julien.corsin@curtin.edu.au}
    \author[2]{Lorenzo Zino} \ead{lorenzo.zino@polito.it}
    \author[1,cor1]{Mengbin Ye} \ead{mengbin.ye@curtin.edu.au}

    \affiliation[1]{organisation = {Centre for Optimisation and Decision Science, Curtin University},
                    city = {Perth WA},
                    country = {Australia}}

    \affiliation[2]{organisation = {Department of Electronics and Telecommunications, Politecnico di Torino},
                    addressline = {Corso Duca degli Abruzzi 24}, 
                    city = {Torino},
                    postcode = {10129},
                    country = {Italy}}

    \cortext[cor1]{Corresponding author: M. Ye. The work of M. Ye is funded by the Western Australian Government through the Premier's Science Fellowship Program, and by the Australian Government through the Office of National Intelligence (NI240100203).}

    \begin{abstract}

        In this paper, we propose an agent-based model of information spread, grounded on psychological insights on the formation and spread of beliefs. In our model, we consider a network of individuals who share two opposing types of information on a specific topic (e.g., pro- vs. anti-vaccine stances), and the accumulation of evidence supporting either type of information is modelled by means of a drift-diffusion process.
        After formalising the model, we put forward a campaign of Monte Carlo simulations to identify population-wide behaviours emerging from agents' exposure to different sources of information, investigating the impact of the number and persistence of such sources, and the role of the network structure through which the individuals interact.
        We find similar emergent behaviours for all network structures considered. When there is a single type of information, the main observed emergent behaviour is consensus. When there are opposing information sources, both consensus or polarisation can result; the latter occurs if the number and persistence of the sources exceeds some threshold values. Importantly, we find the emergent behaviour is mainly influenced by how long the information sources are present for, as opposed to how many sources there are.

    \end{abstract}


    \maketitle

    \section{Introduction}\label{sc:introduction}

        Processes of information spread within social networks have been increasingly and extensively studied over the past few decades \cite{VBZ2016,Alm2019,ZS2021,Mohammadi2024,Zhang2023}. With online communications now accounting for more human interactions than in-person interactions \cite{LS2020}, and considering the ability for false or misleading information (mis/disinformation) to damage health \cite{CSG2008}, democratic \cite{BL2018}, and economic \cite{KMN2018} institutions alike, understanding how ideas spread online has now become an undeniably exigent research problem --- one that is being addressed by the joint efforts of multiple scientific fields.

        At its roots, the science of misinformation is much of a psychological one, with human cognition and social behaviours dominating the causes of false belief formation and adoption \cite{ELC2022}. However, because the spread of information involves countless interactions within large populations, a mathematical abstraction of choice for its study lies in agent-based models (ABMs). In ABMs, individuals (or agents) are arranged according to a given structure (or network topology), and repeatedly interact with one another, updating their states according to mathematical rules \cite{MW2002}. Because of their scalability, ABMs are proficient at bridging the gap between micro- and macro-scale phenomena: in the case of information and misinformation spread, they play a crucial role in shedding light onto how certain population-wide behaviours (e.g. information cascades) emerge from single individuals following cognitive and social processes (e.g. confirmation bias, sharing homogeneity) \cite{VBZ2016,AOP2010,AOS2023,DPZ2020}. While such processes are often well-studied in experimental psychology, their consequences are typically hard to evaluate on larger populations due to experimental constraints. As of today, the literature on modelling misinformation spread using ABMs has at least two central streams.

        One natural approach is to model the spread of misinformation as one of a disease --- a parallel made even more relevant by the COVID-19 crisis recently giving rise to both biological and informational epidemics, termed infodemic~\cite{CQG2020,Bulai2024}. In such ABMs, when an agent has multiple network neighbours, the influence of each connection is independent. Multiple interactions, and interactions repeated over time, increase the probability of information spreading in a linear fashion, though a single interaction suffices for information to spread. Because of this property, such models are called simple contagion models \cite{NMBM2007,TRFM2015}. In contrast, it was also shown that not all spreading processes behave according to the simple contagion paradigm \cite{DCen2007,Watts2007}. In particular, some processes of social diffusion require an individual to be simultaneously exposed to different sources in order for said individual to adopt the contagion, which has been successfully corroborated by empirical evidence \cite{MSFL2017,SH2017}. Since then, so-called complex contagion models have blossomed \cite{GBC2018}, where an agent must be influenced by \emph{multiple} neighbours simultaneously in order for the spreading process to occur, and where the impact of each peer is thus nonlinear.

        The approaches described in the previous paragraph, however, typically assume that the decision-making processes are affected by the current interactions established by the individuals, while the influence of past interactions is typically overlooked. So far, this notion has been explored by few works in related modelling domains, such as decision-making and geopolitics, where we find instances of past-influenced diffusion models such as the social drift-diffusion model (DDM) \cite{TPK2020,FNSS2020} and sandpile models \cite{LYLZ2021}. For this reason, we propose to introduce a new model of information spread that accounts for information accumulation, and to inform its design using insights from cognitive and social psychology, specifically pertaining to memory and false belief formation.

        As pointed out by psychological research, belief formation relies on both past cognition and perceived social consensus \cite{ELC2022,LES2012}. More precisely, it has been determined that upon being presented with a novel piece of information, individuals tend to craft their own belief by assessing that information's compatibility with previously encountered evidence, as well as beliefs of other individuals across their social circle. Possibly because of this interplay, it has been shown that repeated exposure to a given piece of information (even from a single source) suffices to increase the probability of trusting it \cite{HGT1977,BAA2011}. Moreover, it has been observed that false beliefs are able to interfere with an individual's decision-making process for relatively long durations after initial exposure, which has come to be known as the so-called continued influence effect (CIE). Initially formalised in cognitive psychology \cite{JS1994}, this phenomenon has since been extensively studied, with a focus on possible causes as well as mitigation strategies -- both qualitative \cite{KWSO2014,CLE2017} and quantitative \cite{ELSC2011}.

        In order to capture these psychological insights, we endow agents with memory, so that they may hold and accumulate evidence towards belief in one piece of information or another. Then, we place agents on a connected graph, and give them the ability to share this information, based on a combination of both their current state as well as that of neighbouring agents. Finally, in order to model mental corrections and memory alterations, we introduce evidence noise and decay within each agent's internal dynamics. All of these aspects of our model are formally described in \scref{modelspecs}.

        Having introduced these new agent dynamics, we run numerical simulations aiming to model simple scenarios of information spread. Namely, the introduction of single, and later competing, sources of information into an initially at-rest network. Concretely, we model these sources as a subset of agents (which we call committed agents) constantly sharing the same type of information for a set duration. In order to study the impact of committed agents, we also define metrics computed from agent states, which help outline various possible emergent behaviours such as network consensus, and polarisation. Finally, we test the robustness of our conclusions by varying the underlying network topology used to arrange agents and determine their communication capabilities. To this end, we use a few well-established network science models with varied degree distributions~\cite{Boccaletti2006} such as Random-Regular, Watts-Strogatz \cite{WS1998} and Barabási-Albert \cite{AB2002} models, as well as various combinations thereof in the form of Stochastic Block Models.

        For all of these network topologies, we find that our model produces the same set of emergent behaviours. In the absence of committed agents, or when a small number of committed agents are active for a short duration, consensus is the baseline emergent behaviour. Past a certain threshold in both the duration of commitments and number of committed agents, we find that the emerging behaviour depends on how many information types are concurrently introduced in the network. If committed agents of only one type are introduced, consensus is still observed, and in fact occurs faster and more frequently than in the baseline case. However, if two types of committed agents are present and contributing opposite types of information, polarisation emerges. Concluding our findings, we note that, between the number of committed agents and the duration of their commitments, the latter quantity has a stronger influence on which emergent behaviour dominates model simulations. This last result in particular complements the current literature on misinformation spread modelling, wherein quantities analogous to the number of committed agents are often found to more heavily influence the emergent behaviour.

        This article is structured as follows: In \scref{model}, we introduce the evidence-accumulating drift-diffusion model of information spread and motivate its design. Then, in \scref{behaviours}, we use exemplar simulations to identify typical emergent behaviours of the model, and present metrics to quantify the emergence of said behaviours. In \scref{sim_setup}, we describe different scenarios of information spread, and discuss the simulation setup. Finally, \scref{results} presents the main results of the analysis of these scenarios, and \scref{discussion} summarises the main findings and introduces directions for future work.

    \section{Model}\label{sc:model}

        In this section, we formally introduce our evidence-accumulating model for information spread. We start by presenting the population environment and how we characterise the spread of information by means of two coupled variables per agent. Then, we introduce the dynamics that govern the evolution of these two variables. Finally, we discuss our model design and its relation with the existing literature.

        \subsection{Population Environment}\label{sc:modelspecs}

            Consider a population $\mathcal{V} = \{1,\dots,N\}$ of $N$ individuals (agents) who interact on an undirected, unweighted network structure $\mathcal{G} = (\mathcal{V},\mathcal{E})$, where the unordered edge set $\mathcal{E} \subseteq \mathcal{V}\times\mathcal{V}$ represents communication pathways between pairs of agents. Specifically, there is an edge $(i,j) \in \mathcal{E}$ if and only if agents $i$ and $j$ can communicate. We do not consider self-loops, i.e., $(i, i) \notin \mathcal{E}$. Accordingly, we define agent $i$'s neighbourhood $\mathcal{N}_i$ as the set of agents with whom $i$ can communicate. That is, $\mathcal{N}_i = \{j \in \mathcal{V} : (i,j) \in \mathcal{E}\}$.

            Each agent $i \in \mathcal{V}$ is characterised by two interdependent variables $s_i(t) \in \{-1, 0 ,+1\}$ and $c_i(t) \in \mathbb R$, both evolving in discrete time $t \in \NN^* := \{1, 2, \dots\}$.

            The first variable $s_i(t)$ represents the stance of the piece of information being shared by agent~$i$ at time $t$. Specifically, we use $-1$ and $+1$ to represent the orientation of the information that $i$ shares, while $0$ represents the absence of any information sharing. Our model setting considers information sharing on a single topic, where agents are able to share information of opposite stances, e.g. pro/anti-vaccine narratives, left/right political agendas). Thus, $+1$ might represent information supporting a pro-vaccine narrative, in which case $-1$ would then represent information supporting an anti-vaccine narrative. In other words, $s_i(t)$ determines \emph{what} agent $i$ shares. For convenience, we will refer to $s_i(t)$ as the \emph{stance} of agent $i$, even though, as discussed in the above, it more appropriately represents the stance of the information shared, since people do not always share information that is aligned with their internal beliefs.

            The second variable $c_i(t)$ quantifies the amount of evidence that agent~$i$ has accumulated in favour of a particular stance. Negative values of $c_i(t)$ denote accumulating evidence in support of information with stance $-1$, whereas positive values indicate evidence in support of a $+1$-stance. The absolute value of the variable captures the amount of evidence accumulated in favour of the stance. In other words, $c_i(t)$ determines \emph{how inclined} agent $i$ is to share information of either type. We will refer to this variable as the \emph{confidence} of agent~$i$.

            \subsection{Dynamics}\label{sc:dynamics}

                We now describe the dynamics that govern the evolution and the interplay of the two variables $s_i(t)$ and $c_i(t)$. 
                For every agent $i\in \mathcal{V}$, we define the following confidence update rule, which captures evidence accumulation and decay over time:
                \begin{equation}\label{eq:confidence_update}
                        c_i(t + 1) = c_i(t) + \sum_{j\in\mathcal{N}_i} s_j(t) - \delta (c_i(t) - \beta_i) + e_i(t).
                \end{equation}
                Here, $\delta \in [0,1]$ and $\beta_i\in \mathbb R$ are constant parameters and $e_i(t)$ is a sequence of independent and identically distributed Gaussian random variables with mean $0$ and standard deviation $\sigma > 0$, independent for each agent $i \in \mathcal{V}$.

                Equation~(\ref{eq:confidence_update}) poses that, at each time step $t$, agent $i$ revises their confidence $c_i(t)$ according to three distinct mechanisms. Besides the current confidence $c_i(t)$, the first contribution to agent $i$'s revised confidence $c_i(t+1)$ is the term $\sum_{j\in\mathcal{N}_i} s_j(t)$, which accounts for the effect of information communicated by $i$'s neighbours on $i$'s confidence. Namely, agent $i$'s confidence is shifted in the direction of the information received from their social contacts: receiving $s_j(t) = +1$ increases agent $i$'s confidence in the $+1$ stance (and thus $c_i(t)$ increases), while receiving $s_j(t) = -1$ increases agent $i$'s confidence in the $-1$ stance (so, $c_i(t)$ decreases). The second contribution to $c_i(t+1)$ is the term $\delta(c_i(t) - \beta_i)$, which captures the natural decay of agent $i$'s confidence in any stance in the absence of evidence. Specifically, the parameter $\delta \in \mathbb [0,1]$ controls how fast confidence drifts back to the baseline value $\beta_i \in \mathbb R$, which can be interpreted as agent's $i$ personal bias towards a given stance. The third and final contribution to $c_i(t+1)$ is $e_i(t)$, which is a noise term that models the imperfect nature of the information encoding process in human cognition.

                After revising their confidence, each agent re-evaluates whether they wish to share information, and if so, what kind of information. Specifically, agents obey the following stance update rule:
                \begin{equation}\label{eq:stance_update}
                    s_i(t+1) = 
                        \begin{cases}
                            +1 & \text{if } c_i(t+1) > \rho, \\
                            -1 & \text{if } c_i(t+1) < -\rho, \\
                            0 & \text{otherwise},
                        \end{cases}
                \end{equation}
                where $\rho > 0$ is a fixed parameter. In other words, agent $i$ shares information if and only if the magnitude of their confidence exceeds a threshold of $\rho$. In such a case, the shared value is simply the sign of the agent's confidence, as illustrated in the schematic in \fgref{schematic}.

                \begin{figure}
                    \resizebox{\textwidth}{.35\textwidth}{\tikzsetnextfilename{schematic} \begin{tikzpicture}

\tikzstyle{peers} = [draw, circle, text=black, fill=red!40, inner sep = 0pt, minimum size = .7cm]

\node[peers,fill = gray] (1) at (-1.2,3.3) {\color{white}$j$};
\node[peers,fill = gray] (2) at (0.6,2.2) {\color{white}$k$};
\node[peers,fill = gray] (3) at (-0.4,5.4) {\color{white}$h$};
\node[peers,fill = white, minimum size = 1cm] (me) at (1.8,3.8) {$i$};

\path[thick,->] (1) edge[bend right=0] node[above,fill=none] {$s_j$} (me);
\path[thick,->] (2) edge[bend right=0] node[left,fill=none] {$s_k$} (me);
\path[thick,->] (3) edge[bend right=0] node[above,fill=none] {$s_h$} (me);

\draw[-,thick,gray,dashed] (2.5,4.3) edge (4,5.3);
\draw[-,thick,gray,dashed] (2.5,3.3) edge (4,2.3);

\begin{axis}[%
    axis lines = center,
    x axis line style = {->},
    y axis line style = {<->},
    ylabel style = {fill = none, at = {(-.15, 1)}},
    xlabel style = {fill = none, at = {(1, .55)}, anchor = {east}},
    width = 5cm,
    height = 3cm,
    at={(4.5cm, 2.3cm)},
    scale only axis,
    xmin = 0,
    xmax = 9.9,
    ticks = none,
    ylabel = {\footnotesize $c_i(t)$},
    xlabel = {\footnotesize time},
    ymin = -10.5,
    ymax = 10.5,
    clip = false
]

\node[fill = none, color = positivestance, anchor = east] at (0, 5) {\footnotesize $+\rho$};
\node[fill = none, color = negativestance, anchor = east] at (0, -5) {\footnotesize $-\rho$};

\addplot [color = positivestance, very thick, dashed, name path global = A]
    table[row sep=crcr]{
    0	5\\
    10	5\\
};

\addplot [color = negativestance, dashed, very thick, name path global = B]
    table[row sep=crcr]{
    0	-5\\
    10	-5\\
};

\path[name path = ax] (0,10) -- (10,10);
\path[name path = bx] (0,-10) -- (10,-10);

\addplot[positivestance!60] fill between[of = ax and A, soft clip = {domain = 0:10}];
\addplot[negativestance!60] fill between[of = B and bx, soft clip = {domain = 0:10}];

\node[fill = none, anchor = west] at (0 ,7.5) {\color{white}$s_i=+1$};
\node[fill = none, anchor = west] at (0, -7.5) {\color{white}$s_i=-1$};
\node[fill = none, anchor = west] at (0, -3) {\color{gray}$s_i=0$};

\addplot [color=black, very thick, forget plot]
    table[row sep=crcr]{
    0	0\\
    0.05	0.494782516126245\\
    0.1	1.1263292045491\\
    0.15	0.837649721730111\\
    0.2	0.819807196585639\\
    0.25	0.708623759808868\\
    0.3	0.375793216775177\\
    0.35	0.481648098779685\\
    0.4	0.630969804869215\\
    0.45	0.424812243478872\\
    0.5	0.460992207578732\\
    0.55	1.76778309617844\\
    0.6	1.73400005845617\\
    0.65	2.05915152678686\\
    0.7	1.79788184538042\\
    0.75	1.50394289508036\\
    0.8	1.30104936162811\\
    0.85	1.55332824849529\\
    0.9	1.12255867649426\\
    0.95	0.166085283910143\\
    1	0.162693292636736\\
    1.05	-0.0855143158907078\\
    1.1	0.0244894202423739\\
    1.15	0.0367923638328478\\
    1.2	0.371351893616918\\
    1.25	0.791728441439832\\
    1.3	0.0989092941739026\\
    1.35	-0.693380120068603\\
    1.4	-0.472163453537173\\
    1.45	-0.348967740712423\\
    1.5	-0.848547973371754\\
    1.55	-0.25984691015668\\
    1.6	-0.273340952879164\\
    1.65	-0.241556793768549\\
    1.7	-0.134587103993166\\
    1.75	0.212495800148039\\
    1.8	0.944744553014937\\
    1.85	0.783756671887084\\
    1.9	-0.111585081999389\\
    1.95	0.365147067995652\\
    2	-0.128720981227498\\
    2.05	0.463119161884293\\
    2.1	0.243743621472505\\
    2.15	1.06704705265105\\
    2.2	0.733387534767405\\
    2.25	0.249754133043129\\
    2.3	0.805726375762386\\
    2.35	0.196143317536117\\
    2.4	0.61447436572283\\
    2.45	0.907241433348466\\
    2.5	0.906535245678961\\
    2.55	0.806703640921893\\
    2.6	1.90754192154477\\
    2.65	1.78119898461642\\
    2.7	1.74828539604384\\
    2.75	1.34143633677755\\
    2.8	1.00129079121819\\
    2.85	0.77946850754171\\
    2.9	1.70643650772823\\
    2.95	1.57416223376543\\
    3	1.29338792978787\\
    3.05	1.13634369952545\\
    3.1	1.56216843758\\
    3.15	0.968681129062644\\
    3.2	1.35255536540359\\
    3.25	0.753382390119216\\
    3.3	0.760424490338191\\
    3.35	-0.0806278399241739\\
    3.4	0.268139671030145\\
    3.45	0.691890632038163\\
    3.5	0.42119684817741\\
    3.55	0.655772757915848\\
    3.6	0.810636200398824\\
    3.65	1.46208779131016\\
    3.7	1.59166094272886\\
    3.75	2.17547339995243\\
    3.8	2.17533309530064\\
    3.85	1.90943130929551\\
    3.9	1.95134639375873\\
    3.95	1.91217052043509\\
    4	2.41388375123859\\
    4.05	2.44167287827273\\
    4.1	2.07663671479631\\
    4.15	2.05844842860532\\
    4.2	2.7121612795514\\
    4.25	2.55758602713797\\
    4.3	2.28270178561089\\
    4.35	2.83187010180411\\
    4.4	2.93416514806535\\
    4.45	2.80347058452462\\
    4.5	2.48602196276682\\
    4.55	1.80124053802418\\
    4.6	1.93991400738013\\
    4.65	1.84200613137584\\
    4.7	2.06304519626361\\
    4.75	1.39414145032754\\
    4.8	2.12111728676666\\
    4.85	2.42350808267454\\
    4.9	1.69595987727451\\
    4.95	2.28753009297652\\
    5	2.05086589375711\\
    5.05	2.26146804378573\\
    5.1	2.42858455755837\\
    5.15	2.04519816122661\\
    5.2	0.674046299978924\\
    5.25	1.01508348527176\\
    5.3	0.913913995468034\\
    5.35	0.685845428145333\\
    5.4	0.968443328069031\\
    5.45	0.693918534401486\\
    5.5	1.10257486970294\\
    5.55	1.24699785126542\\
    5.6	0.8188684303352\\
    5.65	1.30909471990601\\
    5.7	1.61325825782459\\
    5.75	2.3032086026313\\
    5.8	1.91386518877466\\
    5.85	1.52708942220273\\
    5.9	1.74848149174375\\
    5.95	2.14451325746075\\
    6	2.61712640103791\\
    6.05	2.67420977614119\\
    6.1	2.60952157981142\\
    6.15	2.98983691148973\\
    6.2	4.0222047981554\\
    6.25	4.50427483691173\\
    6.3	4.58150056708069\\
    6.35	4.32842853078064\\
    6.4	4.61501956084453\\
    6.45	4.40681712393016\\
    6.5	4.56197177051512\\
    6.55	4.98547614854744\\
    6.6	5.47775785300991\\
    6.65	5.7970492316266\\
    6.7	6.50804587791263\\
    6.75	7.44856788553199\\
    6.8	6.50524982295433\\
    6.85	5.63085523517088\\
    6.9	5.19322606451581\\
    6.95	4.22233026205687\\
    7	4.08155693057496\\
    7.05	3.9525409813248\\
    7.1	4.11822857390517\\
    7.15	4.30012723875121\\
    7.2	3.99404168230748\\
    7.25	3.62744402161216\\
    7.3	4.06787086082198\\
    7.35	3.91621952523809\\
    7.4	3.27145524490422\\
    7.45	3.32389441966795\\
    7.5	3.47272149834526\\
    7.55	3.31248153019954\\
    7.6	3.6911373045697\\
    7.65	3.84711700124122\\
    7.7	3.78222427466302\\
    7.75	3.80238474641022\\
    7.8	4.07345181037448\\
    7.85	3.89200860286652\\
    7.9	3.40561515569918\\
    7.95	3.37376687438928\\
    8	3.48460313727138\\
    8.05	3.78447126969744\\
    8.1	3.8115647922533\\
    8.15	3.62476666949075\\
    8.2	3.96668681567908\\
    8.25	3.95025572467797\\
    8.3	4.39426537515174\\
    8.35	4.38542135160954\\
    8.4	4.41842107548742\\
    8.45	4.4083065290116\\
    8.5	3.98360432262992\\
    8.55	3.56029251278742\\
    8.6	3.12407925128348\\
    8.65	4.06525337820406\\
    8.7	4.00450635131388\\
    8.75	4.51466711494387\\
    8.8	4.72409861237214\\
    8.85	5.68649777659164\\
    8.9	6.17316878540658\\
    8.95	5.57795517564465\\
    9	5.4385560536411\\
    9.05	5.90999369345869\\
    9.1	6.07498654558936\\
    9.15	7.22683439691963\\
    9.2	7.21062088771855\\
    9.25	6.44258264457525\\
    9.3	6.68019847414531\\
    9.35	6.23748705099909\\
    9.4	6.27982950425846\\
    9.45	6.83605414662556\\
    9.5	7.00178298514986\\
    9.55	7.16158442664942\\
    9.6	7.07135935461203\\
    9.65	7.00743454459449\\
    9.7	6.96582476425909\\
    9.75	5.57667554370029\\
    9.8	5.79754637628596\\
    9.85	6.31469448884904\\
    9.9	5.68273108862454\\
    9.95	5.14491959671959\\
    10	5.28597605458433\\
};

\end{axis}
\end{tikzpicture}}
                    \caption{Schematic of the agents' dynamics. }\label{fg:schematic}
                \end{figure}
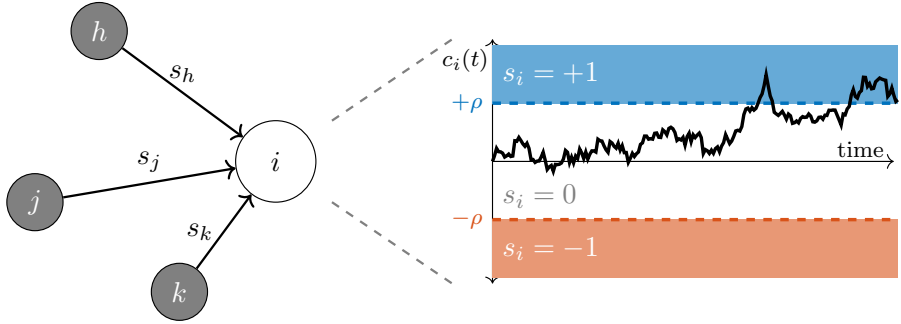

                Observe that the two variables characterising the state of each agent co-evolve and are deeply intertwined. In fact, an agent's confidence depends on the stances received from neighbouring agents through the term $\sum_{j\in\mathcal{N}_i} s_j(t)$, while their own stance is, ultimately, a quantisation of their own confidence. All model variables and parameters associated with an agent are summarised in Table~\ref{tb:parameters}.

                \begin{table}
                    \centering
                    \begin{tabular}{r|l}
                        Symbol & Meaning\\
                        \hline
                        $s_i(t)$ & Stance of individual $i$ at time $t\in\mathbb N$\\
                        $c_i(t)$ & Confidence of individual $i$ at time $t\in\mathbb N$\\
                        \hline
                        $\beta_i$ & Personal bias of individual $i$\\
                        $\delta$ & Confidence decay rate\\
                        $\sigma $ & Standard deviation of confidence noise\\
                        $\rho$ & Sharing threshold\\
                    \end{tabular}
                    \caption{Variables and parameters of the model.}
                    \label{tb:parameters}
                \end{table}

            \subsection{Discussion on the model and related literature}

                The core components of our model design are that of evidence accumulation (captured by $\sum_{j\in\mathcal{N}_i} s_j(t)$), and confidence decay (captured by $\delta(c_i(t) - \beta_i)$), both of which guide agent $i$'s information sharing process. Since information sharing is, at its core, a decision-making problem -- where the decision in question pertains to both \emph{whether} and \emph{what} to share, we motivate our modelling choice by invoking the psychological accounts of decision-making and belief formation.

                First, we observe that the role of evidence accumulation in decision-making has previously been studied by a lineage of cognitive psychology works \cite{LC2004,UTYL2013}, as well as extensively modelled \cite{UM2001,BH2008}. In particular, a tool of choice for the study of evidence accumulation in decision-making is the drift-diffusion model (or DDM) \cite{RM2008}, where evidence takes the shape of external stimuli continuously providing individuals with data supporting one of several alternatives. This accumulation of evidence continues until the amount of evidence for one of the alternatives crosses a threshold, thus completing the decision-making process by selecting said alternative. In our model, \eqref{confidence_update} is based on one particular implementation of the DDM~\cite{TPK2020}, where evidence accumulates not only via external stimuli, but also via \emph{social influence}, a process which has been shown to shape opinion formation processes in social networks \cite{AOP2010}, leading one to conjecture that social drift-diffusion models could be applied to (mis)information spread modelling. Indeed, such a paradigm has been recently used to model the spread of information~\cite{AGG2024}. However, this work differs from ours. In our model, agents share ternary values according to a quantised drift-diffusion process. In \cite{AGG2024}, a DDM is fitted to real-world experimental data so as to derive a reproduction number for a simple contagion model, which predicts the evolution of a binary set of states (share/not share). Ultimately, it is this simple contagion model (an instance of a Susceptible-Infected-Removed epidemic model), rather than a DDM, which is applied to model a process of information spread across the entire network.

                Second, we note that the pervasiveness of misinformation has also been attributed in large part to its persistence in memory, as well as resistance to corrections \cite{LES2012}. This in turn means that even retracted (mis)information can have a continued influence on the decision-making of individuals \cite{JS1994}. In fact, some researchers suspect that conflicting narratives can actually coexist in memory with different strengths of representation \cite{ELC2022} -- which is precisely what the sign and magnitude of $c_i(t)$ model. Consequently, modelling the varying persistence of information representations in memory, we introduce a gradual decay of evidence at a rate $\delta$, towards a baseline evidence level $\beta_i$. This way, varying the strength of $\delta$ is conceptually equivalent to changing how long new evidence influences the decision-making process of agents.

        \section{System's emergent behaviours}\label{sc:behaviours}

            We start presenting our results by illustrating the range of real-world emergent behaviours associated with the spread of information and misinformation that our model is able to reproduce.         
            Specifically, we begin by simulating the model dynamics for a fixed set of model parameters and initial conditions; the precise values of which are not important here, as we focus on qualitative behaviours in order to understand the different emergent behaviours of the model. To this end, different simulation instances of the same set of parameters and initial conditions are shown in \fgref{RR_modes}, illustrating four key emergent behaviours that are typical of this system.

            \begin{figure}
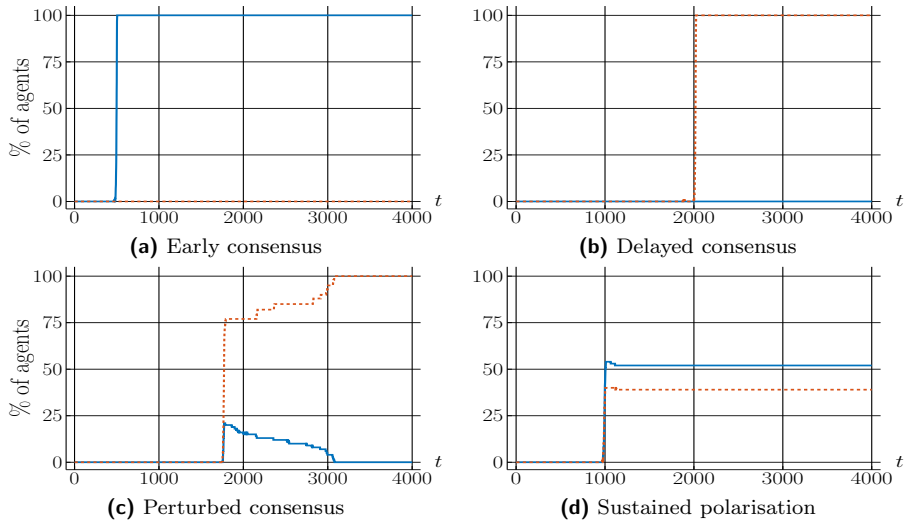


                \begin{subfigure}[t]{.5\textwidth}

                    \resizebox{\textwidth}{.5\textwidth}{\tikzsetnextfilename{RR_nocommit_early_consensus} \input{RR_nocommit_early_consensus.tex}}\\
                    \caption{Early consensus}
                    \label{fg:RR_nocommit_early_consensus}

                \end{subfigure}~\begin{subfigure}[t]{.5\textwidth}
                    \resizebox{\textwidth}{.5\textwidth}{\tikzsetnextfilename{RR_nocommit_delayed_consensus} \input{RR_nocommit_delayed_consensus.tex}}\\
                    \caption{Delayed consensus}
                    \label{fg:RR_nocommit_delayed_consensus}
                \end{subfigure}

                \begin{subfigure}[t]{.5\textwidth}

                    \resizebox{\textwidth}{.5\textwidth}{\tikzsetnextfilename{RR_nocommit_perturbed_consensus} \input{RR_nocommit_perturbed_consensus.tex}}\\
                    \caption{Perturbed consensus}
                    \label{fg:RR_nocommit_perturbed_consensus}

                \end{subfigure}~\begin{subfigure}[t]{.5\textwidth}

                    \resizebox{\textwidth}{.5\textwidth}{\tikzsetnextfilename{RR_nocommit_polarisation} \input{RR_nocommit_polarisation.tex}}\\
                    \caption{Sustained polarisation}
                    \label{fg:RR_nocommit_polarisation}

                \end{subfigure}

                    \caption{Temporal evolution of the fraction of agents sharing $+1$ (solid blue) and $-1$ (dashed orange) for different simulations with the same set of parameters and initial conditions.}
                    \label{fg:RR_modes}

            \end{figure}

            In \fgref{RR_nocommit_early_consensus}, we observe an initial short period where almost no information is shared, after which almost all agents continuously adopt the same stance ($+1$ in this example). This transition from an all-zero to an all-one stance landscape being quite sudden and short-lived ($\approx 30$ time steps in this example) draws a natural parallel with real-world information cascades, characterised by the viral spread of a contagion through a network \cite{JP2017}. Throughout this article, we will refer to an information cascade occurring whenever there is a set of connected agents, all with equal nonzero stances, and the size of this set grows over time while the set remains connected. In the event that the cascade continues until the whole agent population shares the same stance, we say that (stance) \emph{consensus} has been reached.
            However, as \fgref{RR_nocommit_delayed_consensus} demonstrates, such a consensus can sometimes be reached after a substantial delay ($2000$ steps in this example). For this reason, we further nuance this by distinguishing \emph{early consensus} (\fgref{RR_nocommit_early_consensus}) from \emph{delayed consensus} (\fgref{RR_nocommit_delayed_consensus}). 

            In \fgref{RR_nocommit_perturbed_consensus}, a consensus is also asymptotically reached, but differently from \fgref{RR_nocommit_early_consensus} and \fgref{RR_nocommit_delayed_consensus} in two aspects. First, after agents begin to share (just before $t = 2000$), we see that a nontrivial number of agents have a stance of $+1$ in addition to the majority of agents who have stance $-1$. Second, achieving consensus (to $-1$ in this example) is even slower (after $t > 3000$ in this example) than the delayed consensus featured by \fgref{RR_nocommit_delayed_consensus}. Thinking again in terms of information cascades, one could argue that what is observed here is the presence of two candidate cascades of opposite types ($-1$ vs. $+1$) competing for network dominion, with one of them eventually winning, leading to stance consensus. We term this behaviour \emph{perturbed consensus}, distinguishing it from \emph{delayed consensus} because of its transient dynamics, which feature sizeable proportions of agents with both kinds of nonzero stances.
            However, information cascades of opposite types do not necessarily result in one type of information going extinct: in \fgref{RR_nocommit_polarisation}, there is again a sudden shift from no information sharing to agents sharing both kinds of information (at $t = 1000$ in this example), but this situation remains stable over the rest of the simulation time window. Because it is reminiscent of situations of social diffusion where two alternatives coexist in a network, we shall refer to this emergent behaviour as \emph{sustained polarisation}.

            \subsection{Metrics to classify the emergent behaviours}\label{sc:metrics}

                The differences between the five emergent behaviours identified above can easily be observed in a qualitative manner, as was done by plotting the stance profiles of various simulations in \fgref{RR_modes}. However, given the stochastic nature of the dynamics, different behaviours can emerge from using the same model parameters (as exemplified by all plots from \fgref{RR_modes} being generated with the same set of model parameters). To account for this when quantifying the emergence of given behaviours, the main results of this paper are obtained via a large number of Monte Carlo simulations. Because of this, we develop a systematic and quantitative approach to differentiate the aforementioned emergent behaviours, motivating us to now introduce a set of population-level metrics. 

                In what follows, consider a population of $N \in \NN$ agents, whose states obey the dynamics governed by \eqref{confidence_update} and \eqref{stance_update} for a fixed number of $T \in \NN$ time steps.

                For each agent $i\in \mathcal{V}$, define a three-dimensional vector $x^i = [x^i_1,x^i_2,x^i_3]$, with $x^i_1 := \frac1T|\{t \leq T : s_i(t)=+1\}|$, $x^i_2 := \frac1T|\{t \leq T : s_i(t)=-1\}|$, and $x^i_3 := \frac1T|\{t \leq T : s_i(t)=0\}|$.
                In other words, this vector measures the respective \emph{fractions of time for which agent $i$} shared pieces of information of either type, or did not share. Building on this, we define the \emph{time-wise entropy}     of agent $i\in \mathcal{V}$ as:
                \begin{equation}\label{eq:tw_entropy}
                    X(i) := -\sum_{j=1}^3x^{i}_{j}\log_3(x^{i}_{j}),
                \end{equation}
                with the convention that $x\log_3(x) = 0$ for $x = 0$.
    
                Next, for each time $t \leq T$, we define a three-dimensional vector $y^t=[y^t_1,y^t_2,y^t_3]$, with $y^t_1 := \frac1N|\{i\in \mathcal{V} : s_i(t)=+1\}|$, $y^t_2 := |\frac1N\{i\in \mathcal{V} : s_i(t)=-1\}|$, and $y^t_3 := \frac1N|\{i\in \mathcal{V} : s_i(t)=0\}|$. In other words, this vector measures the respective \emph{fractions of agents who at time $t$} shared pieces of information of either type, or did not share. Again, we introduce the entropy of $y^t$, or \emph{agent-wise entropy} at time $t \leq T$:
                \begin{equation}\label{eq:aw_entropy}
                    Y(t) := -\sum_{j=1}^3y^{t}_{j}\log_3(y^{t}_{j}).
                \end{equation}

                Because $x^i$ and $y^t$ are stochastic vectors, $X(i)$ and $Y(t)$ are bounded in the interval $[0,1]$. They are equal to $1$ if stances are uniformly distributed across their respective domains, and to $0$ if all stances are equal over said domains. Therefore, a large value of time-wise entropy $X(i)$ characterises an agent $i$ who frequently changes the stance of the information they share over time, while a large value of agent-wise entropy $Y(t)$ indicates a specific time $t$ for which the stances of the information being shared are quite diverse across agents.

                Furthermore, notice that $X(i)$ and $Y(t)$ are respective functions of a single agent and a single time step. However, to capture emergent behaviour at the population level, it is useful to aggregate these quantities over agent space, and time. For this reason, we define the respective \emph{temporal entropy} and \emph{spatial entropy} of stances as
                \begin{equation*}
                    \bar{X} = \frac{1}{N}\sum_{i = 1}^{N}X(i) \qquad \bar{Y} = \frac{1}{T}\sum_{t = 1}^{T}Y(t).
                \end{equation*}
                In the following, we use $\bar{X}$ and $\bar{Y}$ to characterise each of the aforementioned model behaviours, as summarised by \fgref{ROC}.

                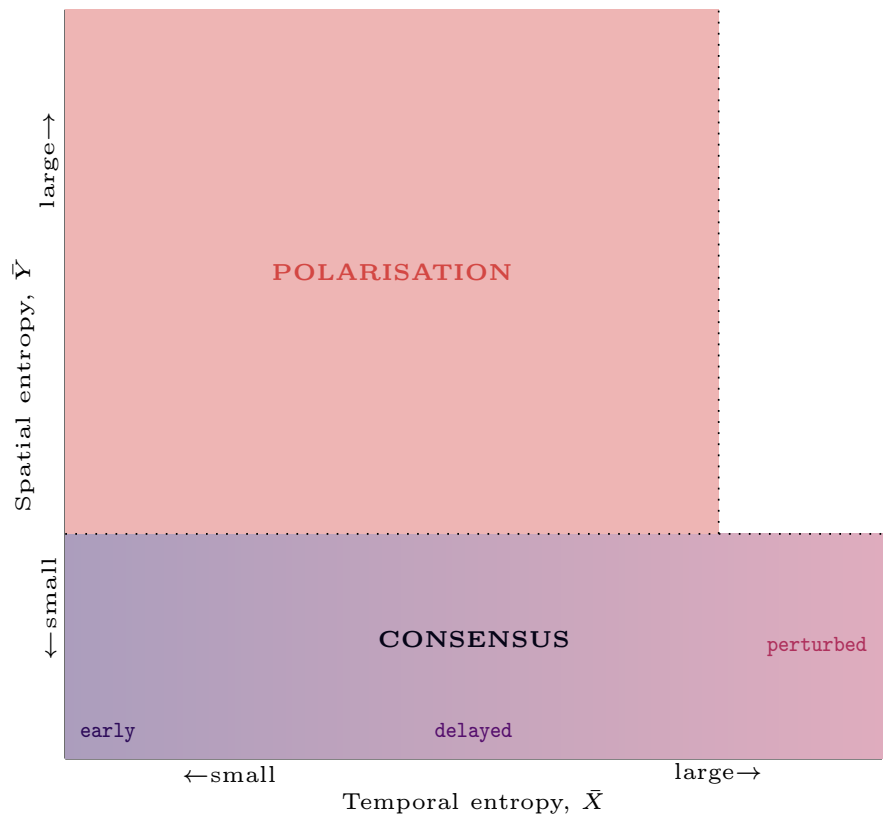
\begin{figure}

                    \centering \resizebox{\textwidth}{.9\textwidth}{\tikzsetnextfilename{ROC} \begin{tikzpicture}

    \definecolor{cons}{rgb}{0.17649300,0.04140200,0.34811100}
    \definecolor{conss}{rgb}{0.33521700,0.06006000,0.42852400}
    \definecolor{consss}{rgb}{0.68265600,0.18950100,0.36075700}
    \definecolor{pol}{rgb}{0.82737200,0.27951700,0.26175000}

    \begin{axis}[clip = false, ticks = none, 
                 xmin = 0, xmax = 1, 
                 xlabel = {\tiny Temporal entropy, $\bar{X}$}, xlabel style = {at = {(.5, -0.02)}}, 
                 axis x line* = left, 
                 x axis line style={-},
                 ymin = 0, ymax = 1, 
                 ylabel = {\tiny Spatial entropy, $\bar{Y}$}, ylabel style = {at = {(-.02, .5)}, rotate = 0}, axis y line* = left,
                 y axis line style={-},]

        \node at (.4, .65) {\tiny \textcolor[rgb]{0.82737200,0.27951700,0.26175000}{\textbf{POLARISATION}}};
        \draw[dotted] (.8, .3) -- (.8, 1);
        \draw[dotted, name path = ax] (0, .3) -- (1, .3);
        \node at (.5, .16) {\tiny \textcolor[rgb]{0.01937300,0.01513300,0.08876700}{\textbf{CONSENSUS}}};
        \node[anchor = east] at (1, .15) {\tiny \textcolor[rgb]{0.68265600,0.18950100,0.36075700}{\texttt{perturbed}}};
        \node[anchor = south] at (.5, 0) {\tiny \textcolor[rgb]{0.33521700,0.06006000,0.42852400}{\texttt{delayed}}};
        \node[anchor = south west] at (.0, 0) {\tiny \textcolor[rgb]{0.17649300,0.04140200,0.34811100}{\texttt{early}}};

        \path[name path = paxis] (axis cs:0,0) -- (axis cs:1,0);
        \path[name path = high] (axis cs:0,1) -- (axis cs:1,1);

        \addplot[left color = cons!40, right color = consss!40] fill between [of = paxis and ax, soft clip = {domain = 0:1}];
        \addplot[pol!40] fill between [of =  ax and high, soft clip = {domain = 0:.8}];

        \node at (0.8, -0.02) {\tiny large$\rightarrow$};
        \node at (0.2, -0.02) {\tiny $\leftarrow$small};

        \node[rotate=90] at (-0.02,.8) {\tiny large$\rightarrow$};
        \node[rotate=90] at (-0.02,.2) {\tiny $\leftarrow$small};

    \end{axis}
\end{tikzpicture}}
                    \caption{Characterisation of possible emergent behaviours of the model as a function of temporal ($\bar{X}$) and spatial ($\bar{Y}$) entropy.
                    }
                    \label{fg:ROC}

                \end{figure}
    
                First, consider small values of spatial entropy $\bar{Y}$: As previously mentioned, this happens when stances do not substantially vary across agents. Aside from the limit scenario where $\bar{X} = \bar{Y} = 0$, this corresponds to stance \emph{consensus} occurring, with $\bar{X}$ quantifying how \emph{delayed}, and $\bar{Y}$ how \emph{perturbed}.
                Second, consider large values of spatial entropy $\bar{Y}$: As previously mentioned, this happens when the agent population is fractured into large groups of differing stances. Should these be stable over time (i.e. should $\bar{X}$ be small enough), then \emph{polarisation} can be said to occur, as agents effectively adopt heterogeneous long-term stances. If not, then both entropies are large, which means that agents stances vary a lot across both agents and time. In this case, there is no easily discernible emergent behaviour at the population level, although this rarely happens for the model parameters studied in this work.

                While conveniently summarising agent stances with one pair of values, $\bar{X}$ and $\bar{Y}$ present a few limitations. First, they quantify stance differences in such a way that the difference between not sharing and sharing is considered equally significant as the one between sharing opposite types of information. Additionally, they do not capture the network topological characteristics of these stance differences, meaning that a situation involving many non-adjacent agents with the same nonzero stances would feature high $\bar{Y}$ and small $\bar{X}$, meaning it would be interpreted as polarisation if inspected only under the lens of entropy alone.

                For these reasons, we introduce a third metric capturing disagreement within the network, which accounts for topological aspects of the model dynamics. Specifically, inspired from the potential function of the coordination games on networks from  \cite{AYZ2023}, we define \emph{network disagreement} at time $t$ as follows:
                \begin{equation}\label{eq:disagreement}
                    D(t) := \frac1{4|\mathcal{E}|}\sum_{i \in \mathcal{V}}\sum_{j \in \mathcal{N}_i}(1 - s_i(t)s_j(t)) = \frac1{2|\mathcal{E}|}\sum_{(i,j) \in \mathcal{E}}(1 - s_i(t)s_j(t)).
                \end{equation}
                In \eqref{disagreement}, the summand $1 - s_i(t)s_j(t)$ measures a notion of \emph{pairwise disagreement} between agents $i$ and $j$. Mathematically, the only values pairwise disagreement can take are $0$, $1$, and $2$, depending on the stances of the agents considered. Specifically, pairwise disagreement is maximal whenever $s_i(t)$ and $s_j(t)$ are nonzero and opposite, minimal when they are nonzero and equal, and takes an intermediate value when either or both are zero. Because of this, agent pairs $(i,j) \in \mathcal{E}$ with $s_i(t) = 0$ and $|s_j(t)| = 1$ are considered less disagreeing than pairs with $s_i(t) = +1$ and $s_j(t) = -1$ or vice-versa (a nuance which was not captured by the entropy metrics).
                Moreover, because pairwise disagreement takes discrete values between $0$ and $2$ for any agent pair and any time step $t \in \NN$, the network disagreement $D(t)$ itself evolves within the interval $[0,1]$ as agents dynamics from \scref{dynamics} play out. In particular, note that all=zero stances, \emph{consensus}, and \emph{polarisation} can once again all be captured, respectively so by the sequence of network disagreement converging to $0.5$, $0$, or a different fixed constant $D^* \in [0, 1]$. In the latter case, this value $D^*$ is then proportional to the number of agents on the boundary of polarised agent clusters. This last property in particular helps differentiate situations of polarisation where same-stance agents are spatially separated by agents of differing stances.
                Finally, for the sake of convenience, we define a simulation-wide summary of network disagreement, by taking the time-average of $D(t)$: 
                \begin{equation*}
                    \bar{D} := \frac{1}{T}\sum_{t = 0}^{T}D(t),
                \end{equation*}
                which completes our metrics definitions.

    \section{Research questions and simulation setup}\label{sc:sim_setup}

        With the model defined in \scref{model} and its emergent behaviours discussed and classified in \scref{behaviours}, we are now ready to present our research questions and describe the methodology used to implement and simulate our model. First, we design two different scenarios of information spread for which we will investigate the system's emergent behaviours. To distinguish these scenarios, we introduce new simulation-specific parameters (meta-parameters in what follows). Second, we discuss the concrete implementation of our simulations and we provide details on the region of the parameter space explored. Third, we formulate our key research questions.

        \subsection{Scenarios of information spread} \label{sc:sim_scenarios}

            In order to understand how our evidence accumulation process shapes the spread of information across a network, we design and run a set of Monte Carlo simulations on various synthetic network topologies. Each set of simulations is designed so as to represent different scenarios of information spread, each differing by i) the varying presence and ii) the persistence of \emph{information sources}. To quantify the former, we introduce the rational numbers $(\zeta^+,\zeta^-) \in [0,1]^2$ representing the fractions of information sources associated with respectively the $+1$ and $-1$ stances with respect to the total population size $N$. To quantify the latter, we introduce the non-negative integer $\tau < T$, which represents how long these sources remain active for. In what follows, we will refer to the simulation parameters $(\tau, \zeta^+, \zeta^-)$ as meta-parameters, allowing us to define each of the information spread scenarios of interest as an exploration of one particular subspace of this three-dimensional meta-parameter space.

            \subsubsection{Scenario I: Single information source}\label{sc:single_source}

                In our first scenario, we model a situation where information of a particular stance is introduced into a community that is otherwise initially inactive; the information being introduced by a small number of committed minority individuals. This allows us to study how information may propagate when introduced for the first time into a network.
                For this scenario, and since the model dynamics are symmetric around $s_i(t) = 0$, we choose -- without loss of generality -- to model committed minority individuals who only share information with stance $+1$. To this end, we set $\zeta^+ > 0$ and $\zeta^- = 0$. Then, we uniformly sample a subset of $\zeta^+N$ committed agents $\mathcal{Q}^+ \subseteq \mathcal{V}$ and force all agents in $\mathcal{Q}^+$ to share $+1$ at every time step $t \leq \tau$, regardless of their confidence $c_i(t)$. Once $t > \tau$, the stances of committed agents resume updating according to the rules described in \scref{dynamics}. For brevity, we may later refer to this scenario as the single-source case.

            \subsubsection{Scenario II: Opposing information sources}\label{sc:opposing_sources}

                Next, we turn to a more complex scenario, where two groups of individuals with opposite stances compete for the public information landscape.
                For this, we follow a similar simulation protocol as was discussed in \scref{single_source}, except for the fact that, besides the group of $\zeta^+N$ committed agents who share information with stance $+1$, there is also a second group of $\zeta^-N$ committed agents who share information with stance $-1$.

                Formally, we implement this scenario by selecting two disjoint subsets of committed agents $\mathcal{Q}^-,\mathcal{Q}^+ \subset \mathcal{V}$ uniformly at random from $\mathcal{V}$ such that $\mathcal{Q}^- \cap \mathcal{Q}^+ = \emptyset$, $|\mathcal{Q}^-| = \zeta^-N$ and $|\mathcal{Q}^+| = \zeta^+ N$. We then set agents in $\mathcal{Q}^-$ to share $-1$ while agents in $\mathcal{Q}^+$ are made to share $+1$ at each time step for all $t \leq \tau$. As above, for $t > \tau$, all committed agents revert to the usual dynamics. For brevity, we may later refer to this scenario as the dual-source case.

        \subsection{Simulation details} \label{sc:sim_process}

            We use Julia \cite{BEKS2017} to implement Monte Carlo simulations of the aforementioned scenarios. The full simulation source code is provided at \cite{Corsin2024} and a high-level description of the corresponding algorithm is provided hereafter. 

            First, the state variables of non-committed agents are initialised as follows:
            \begin{equation*}\label{eq:normal_init}
                \forall i \in \mathcal{V} \setminus (Q^- \cup Q^+) \qquad s_i(0) = 0 \qquad c_i(0) = \beta_i,
            \end{equation*}
            while those of committed agents are forced to $s_i(0) = \pm1$, according to the scenario of information spread considered (see \scref{sim_scenarios}).
            Following this, all non-committed agents update their state variables exactly $T$ times using \eqref{confidence_update} followed by \eqref{stance_update}. Committed agents do not update their stance until $\tau$ steps have passed. At each time step $1 \leq t \leq T$ of the simulation process, we gather the simulated agent stances and confidences, which we use to compute the metrics defined in \scref{metrics}. Finally, in order to obtain valid Monte Carlo approximations of said metrics, each simulation is repeated a total of $R$ times, in order to obtain a sample mean and standard deviation of each metric -- which are the final metric values presented throughout the results in \scref{results}. Since the set of preliminary simulations presented in \fgref{RR_modes} suggests that few as $T = 3000$ time steps are sufficient for any of the emergent behaviours of our model to stabilise, we elect to stop simulations after reaching this many iterations. Additionally, we fix the number of Monte Carlo samples to $R = 1024$, which we found to be sufficient for the Monte Carlo sample standard deviations of all metrics $(\bar{X}, \bar{Y}, \bar{D})$ to be negligible.

            \subsubsection{Parameter space}\label{sc:param_heuristics}

                In~\ref{sc:analysis}, we derive some analytical insights into the model, allowing us to restrict the parameter space to a sub-space of interest for our studies. First, we choose to remove any bias by setting $\beta_i = 0$ for all agents $i \in \mathcal{V}$. In fact, from our analytical observations summarised in \eqref{equilibrium}, varying $\beta_i\neq 0$ would likely have a similar effect as varying $\zeta^+$ and $\zeta^-$ on long-term model dynamics. 

                Second, since the scope of our simulations is to explore the impact of the meta-parameters $\tau$ and $\zeta^+$, we fix the remaining model parameters to values which do not substantially determine the overall emergent behaviours observed. In our analytical computations from \eqref{switchpr_bound}, we observe that to avoid pathological patterns in how individuals share information (e.g. sudden stance transitions from $-1$ to $+1$), we need to enforce the threshold $\rho$ to be larger than agents' degrees. For this reason, if we let $K$ be the average degree of the network, we fix $\rho = K + 1$. Moreover, since we do not want confidence noise to be the main contributor to the formation of information cascades, we fix $\sigma = 0.2$ based on \eqref{switchpr_bound}. This way, the impact of a change to neighbouring stances on an agent's confidence is reasonably unlikely to be eclipsed by noise alone. Finally, we fix $\delta = 0.01$ so that the equilibrium point of agent confidences in \eqref{equilibrium} remains far enough from zero, thus enabling agent stances to remain in a (non-zero) steady state with high probability once consensus is established. All of these adjustments to the parameter space are summarised in \tbref{params_reduced}.

                \begin{table}
                    \begin{center}
                        \begin{tabular}{  p{0.05\linewidth} | >{\centering}p{0.15\linewidth} | p{0.6\linewidth}  }
                                   & Values & Definition \\
                            \hline
                            $\tau$ & $\NN_{\leq T}$ & \textbf{Duration of agent commitment (steps)} \\
                            \hline
                            $\zeta^+$ & $[0,1]$ & \textbf{Fraction of committed agents with stance $+1$} \\
                            \hline
                            $\zeta^-$ & $[0,\zeta^+]$ & Fraction of committed agents with stance $-1$ \\
                            \hline
                            $\rho$ & $K + 1$ & Sharing threshold \\
                            \hline
                            $\delta$ & $0.01$ & \textit{Confidence decay rate} \\
                            \hline
                            $\beta_i$ & $0$ & \textit{Personal bias of individual $i$} \\
                            \hline
                            $\sigma$ & $0.2$ & \textit{Standard deviation of confidence noise} \\ 
                            \hline
                            $N$ & $\{100,1000\}$ & \textit{Number of agents in the network} \\ 
                            \hline
                            $K$ & $4$ & \textit{Average degree of the network} \\ 
                        \end{tabular}
                    \end{center}
                    \caption{\footnotesize Reduced parameter ranges. Bold definitions denote free parameters. Italic definitions denote fixed parameters. Definitions in roman font denote dependent parameters. The last two parameters are associated with the network structure and are changed in the last set of simulations only.}
    
                \label{tb:params_reduced}
                \end{table}

            \subsubsection{Network topologies}\label{sc:other_topologies}

                Our agents interact on a network structure, which determines the set of neighbours from which an agent can receive information (as per \eqref{confidence_update}). For the underlying network topology used in our simulations, we consider four different types of network topologies, briefly described hereafter.

                \paragraph{Random Regular (RR) Networks}\label{pg:RR} In RR networks, the set of agent connections $\mathcal{E}$ is sampled uniformly at random from the set of all possible configurations in which every agent has the same constant number of $K$ neighbours. That is, $|\mathcal{N}_i| = K \in \NN$ for every agent $i \in \mathcal{V}$.

                \paragraph{Watts-Strogatz (WS) Model}\label{pg:WS} Part of a broader category of ``Small-World'' networks, this is obtained by first generating a network where agents are arranged in a circular fashion, each connected to their closest $\frac{K}{2}$ neighbours on each side. Then, for each node, each connection starting from the node is rewired with probability $\eta \in [0,1]$, independently of the others. The rewiring is such that the connection is replaced by another one, starting from the same node but connecting it to another node selected uniformly at random from the remaining $N-2$ nodes in $\mathcal V$. In this work, we consider a rewiring probability of $\eta = 0.1$. This process results in networks with several long-range interactions that reduces the average distance between nodes, yielding the so-called small world behaviours that is typical of many real-world social communities~\cite{Newman2003}.

                \paragraph{Barabási–Albert (BA) Model}\label{pg:BA} Part of a broader category of ``Scale-Free'' networks, these constructions are generated by iteratively adding agents to a fully-connected set of agents according to a preferential attachment rule. Specifically, each new agent chooses $\frac{K}{2}$ existing agents, with probability proportional to the current degree of existing agents, yielding a rich-get-richer phenomenon. This process  results in networks possessing a power-law distribution of degrees -- a property that is also observed (to some extent) in many real-world social networks~\cite{Clauset2009}.

                \paragraph{Stochastic Block Model (SBM)}\label{pg:SBM} This model combines two existing networks (or sub-communities) by pairing agents form either network together with some fixed probability $\lambda$. In this work, we consider combinations of RR and BA networks, and a link probability of $\lambda = \frac{1}{20}(\frac{K}{N} + \frac{K}{N}) = 0.004$, which is negligible with respect to both of these networks' respective average degrees. 

            \subsection{Research questions}\label{sc:sim_questions}

                We start our exploration of the model by fixing the network topology, and studying what behaviours emerge for each scenario of information spread described in \scref{sim_scenarios}, when the model parameters are varied within the ranges of \tbref{params_reduced}. To perform such a study, we use Random-Regular (RR) networks with $N = 100$ agents and $K = 4$. The choice of using RR networks is motivated by two main reasons. First, in RR networks all agents have the same number of neighbours, reducing the potential confounding of the population heterogeneity on the characterisation of the emergent behaviours. Second, the randomness in the network formation process allows us to introduce some randomness in the placement of information sources, providing robustness to our findings. 

                Once baseline results have been established for RR networks, we repeat all our simulations on more complex network topologies in order to examine the role of the network and the robustness of our findings. First, we start by investigating the scalability of the dual-source scenario by re-running the simulations from \scref{opposing_sources} on larger ($N = 1000$) RR networks. Next, we repeat the dual-source scenario on the other synthetic network topologies described above.

    \section{Results}\label{sc:results}

        In this section, we explore the research questions described in \scref{sim_questions} by means of Monte Carlo simulations, as described in \scref{sim_process}. We proceed in the same order as when introducing the different scenarios of information spread. Namely, we first investigate the single-source case, followed by the dual-source case, before studying how our observations for the dual-source case generalise to larger and different network constructions.

        \subsection{Single information source}\label{sc:results_single}

            \begin{figure}
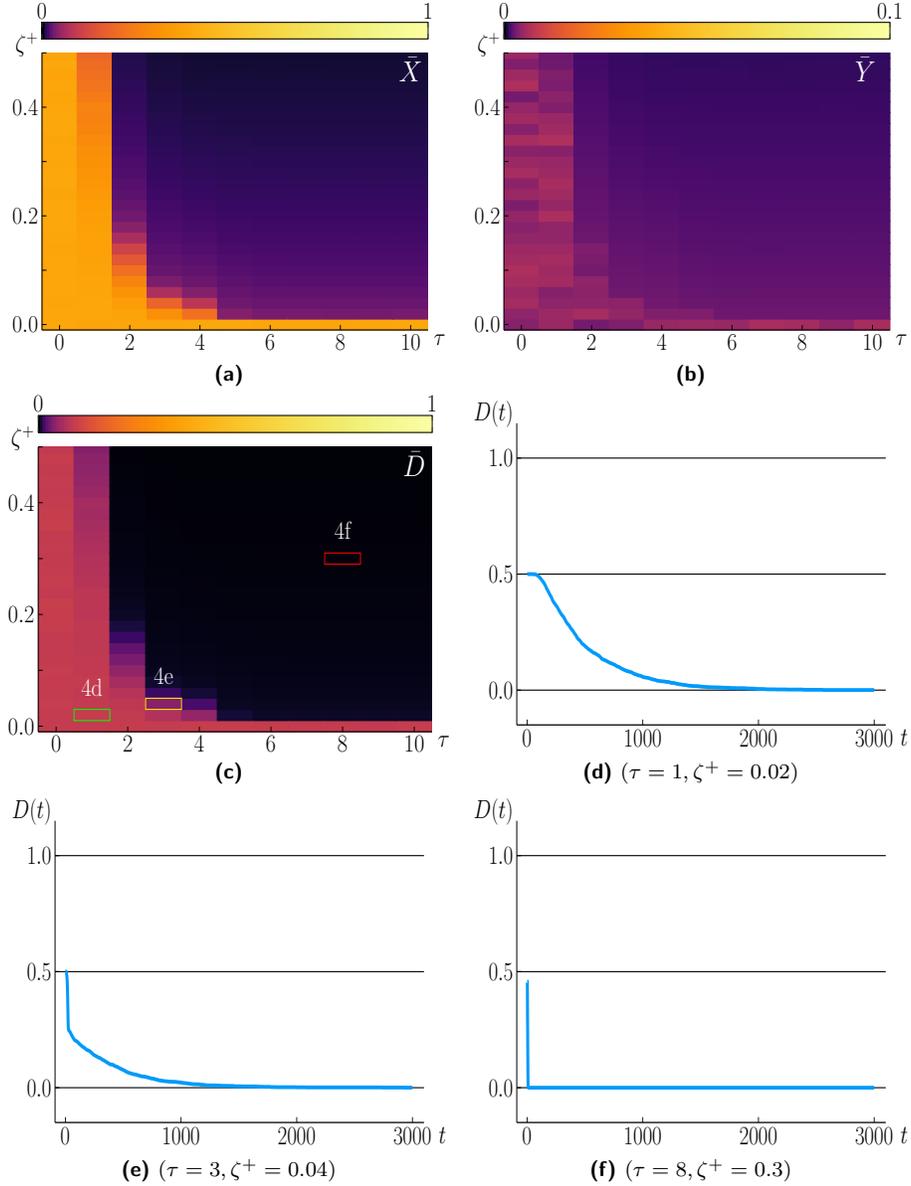

                \begin{subfigure}[t]{.5\textwidth}

                    \resizebox{\textwidth}{.8\textwidth}{\tikzsetnextfilename{DDM_RR_single_Xmap} \input{DDM_RR_single_Xmap.tex}}\\
                    \caption{}
                    \label{fg:DDM_RR_single_Xmap}

                \end{subfigure}~\begin{subfigure}[t]{.5\textwidth}

                    \resizebox{\textwidth}{.8\textwidth}{\tikzsetnextfilename{DDM_RR_single_Ymap} \input{DDM_RR_single_Ymap.tex}}\\
                    \caption{}
                    \label{fg:DDM_RR_single_Ymap}

                \end{subfigure}

                \begin{subfigure}[t]{.5\textwidth}

                    \resizebox{\textwidth}{.8\textwidth}{\tikzsetnextfilename{DDM_RR_single_Dmap} \input{DDM_RR_single_Dmap.tex}}\\
                    \caption{}
                    \label{fg:DDM_RR_single_Dmap}

                \end{subfigure}~\begin{subfigure}[t]{.5\textwidth}

                    \resizebox{\textwidth}{.8\textwidth}{\tikzsetnextfilename{DDM_RR_single_Dt_low} \input{DDM_RR_single_Dt_low.tex}}\\
                    \caption{$(\tau = 1, \zeta^+ = 0.02)$}
                    \label{fg:DDM_RR_single_Dt_low}

                \end{subfigure}

                \begin{subfigure}[t]{.5\textwidth}

                    \resizebox{\textwidth}{.8\textwidth}{\tikzsetnextfilename{DDM_RR_single_Dt_mid} \input{DDM_RR_single_Dt_mid.tex}}\\
                    \caption{$(\tau = 3, \zeta^+ = 0.04)$}
                    \label{fg:DDM_RR_single_Dt_mid}

                \end{subfigure}~\begin{subfigure}[t]{.5\textwidth}

                    \resizebox{\textwidth}{.8\textwidth}{\tikzsetnextfilename{DDM_RR_single_Dt_high} \input{DDM_RR_single_Dt_high.tex}}\\
                    \caption{$(\tau = 8, \zeta^+ = 0.3)$}
                    \label{fg:DDM_RR_single_Dt_high}

                \end{subfigure}

                \caption{Results of the Monte Carlo simulations for a RR network in the single-source case ($\zeta^- = 0$). In panels {\bf(a)}, {\bf(b)}, and {\bf(c)}, we report the values of the temporal entropy $\bar X$, spatial entropy $\bar Y$, and average network disagreement $\bar D$, respectively, using a colour code, for different agent commitment durations $\tau$ and fractions of committed agents $\zeta^+$. In panels {\bf(d--f)}, we represent the temporal evolution of the network disagreement $D(t)$ for three pairs of meta-parameters $(\tau,\zeta^+)$ of interest,  represented by the boxed areas in panel {\bf(c)}.}
                \label{fg:DDM_RR_single}

            \end{figure}

            First, we examine the effect of a committed minority of agents sharing information of the same stance on a RR network, following the simulation setup described in \scref{single_source}. To this end, we turn to \fgref{DDM_RR_single}, which shows measurements of $(\bar{X},\bar{Y},\bar{D}$) for a region of interest of the single-source meta-parameter space, namely the $(\tau, \zeta^+)$ plane. For $\bar{Y}$, note that the colour scale used in \fgref{DDM_RR_single_Ymap} differs from the other panels, so as to display negligible values in a visible way.

            From the heatmaps of temporal entropy $\bar{X}$ (\fgref{DDM_RR_single_Xmap}) and average network disagreement $\bar{D}$ (\fgref{DDM_RR_single_Dmap}), we identify two observations (or sub-scenarios) of interest, conditional on joint values of $\tau$ and $\zeta^+$. First, we discuss the case of \emph{weak agent commitments}, that is, commitments which are short-lived (low $\tau$) and involve few committed agents (low $\zeta^+$). Then, we discuss the case of \emph{strong agent commitments}, i.e. commitments which are either long-lived (high $\tau$) or involve a large fraction of committed agents (high $\zeta^+$). Finally, we conclude our analysis of the single-source scenario by comparing the independent influences of the duration of commitment $\tau$ and the fraction of committed agents $\zeta^+$ to determine which of the two has a stronger impact on the emergent behaviours observed for this scenario.

            \paragraph{Weak agent commitments} When committed agents are few, and their commitments short-lived, most simulations yield delayed consensus, though we also observe a few instances of perturbed consensus. The first of these observations can be inferred by looking at the region of \fgref{DDM_RR_single_Dmap} and \fgref{DDM_RR_single_Xmap} with $\tau < 5$, which respectively feature a low average network disagreement ($\bar{D} < 0.2$) and a high temporal entropy ($\bar{X} >> \bar{Y} \simeq 0$); both of which we have previously identified in \scref{metrics} as markers of delayed consensus. The presence of rare instances of perturbed consensus, on the other hand, can be inferred from the negligible but nonzero values of spatial entropy in \fgref{DDM_RR_single_Ymap} for $\tau < 5$, as well as the relatively slow ($t > 1000$) convergence of $D(t)$ to $0$ on \fgref{DDM_RR_single_Dt_low}.

            We conjecture that the delayed consensus observed for weak agent commitments is a consequence of information cascades having a very low probability of forming within the simulation time window. In order to understand why this would be true, consider that in our model, information cascades form if a sufficiently large set of connected agents concurrently maintain their absolute confidences above the sharing threshold $\rho$ (and thus are sharing), for a sufficiently long period to start a feedback loop in \eqref{confidence_update}. Since they both facilitate the occurrence of these conditions, evidence noise and committed agents are both possible motors of cascade formation, and thus consensus. However, observe that delayed consensus is the dominant emergent behaviour in the absence of any committed agents ($\tau = 0$ or $\zeta^+ = 0$) as well as for weak agent commitments. This suggests that the main contributor to an information cascade within the simulation time window is unlikely to be the presence of committed agents, but rather evidence noise (notice that with $\sigma > 0$, a cascade forms almost surely as $t \to \infty$). However, recall that the model parameters were chosen in \scref{param_heuristics} so as to minimise the impact of evidence noise on cascade formation. Consequently, within our simulations, noise-induced information cascades occur with a low probability; expected time to cascade formation, and in turn the average time to consensus, are therefore large.

            Then, to explain why perturbed consensus is sometimes observed, recall that the distribution of evidence noise is centred around $0$. In particular, this means that information cascades associated with both $-1$ and $+1$ stances have a low probability of being induced by evidence noise. In turn, it is unlikely for two information cascades to form simultaneously, making it likely that there is a significant time gap between (noise-induced) information cascades of opposite stances. Because of this, and because of the regularity of RR networks, if both $+1$ and $-1$ cascades form during a simulation, one of them will involve significantly more agents than the other by the time both information cascades meet within the network. However, at the interface of these information cascades, agents are more likely to convert their stance to that of the largest cascade as time progresses (a mathematical intuition for this fact is available under \eqref{equilibrium}). Because of the size difference between the two cascades, this conversion process is unlikely to halt until all agents are eventually of the same stance, thus explaining the observed instances of perturbed consensus.

            \paragraph{Strong agent commitments} Increasing either of the commitments duration, or committed agents fraction, seems to have a two-fold effect on the aforementioned behaviours. First, the time needed to reach consensus is drastically reduced. Second, the previously rare instances of perturbed consensus are now nonexistent. The former effect can be seen in the fast decreasing temporal entropy $\bar{X}$ and average network disagreement $\bar{D}$ of \fgref{DDM_RR_single_Xmap} and \fgref{DDM_RR_single_Dmap} respectively, as the commitment duration increases above $\tau = 5$. The latter effect is characterised by the now-zero spatial entropy $\bar{Y}$ for $\tau > 5$ in \fgref{DDM_RR_single_Ymap}.

            Analogously to the weak agent commitments case, we conjecture that these observations come as a consequence of strong agent commitments yielding a much higher probability of forming a ($+1$) information cascade within the simulation time window --- which in turn yields shorter average times to consensus. Building upon the explanation given for weak agent commitments, we note two straightforward ways to increase the chance of a $+1$ cascade forming. Specifically, one can either reduce the size of the consistently-sharing agent cluster which is needed to initiate a positive confidence feedback loop in \eqref{confidence_update}, or reduce how long this cluster needs to share for. Effectively, strong agent commitments achieve either, for the former is achieved by adding more committed agents, while the latter is achieved by increasing commitment length.

            Finally, since in this scenario, only committed agents with stance $+1$ are introduced, the probability of information cascades of stance $-1$ forming does not increase when varying either meta-parameter. Consequently, in this sub-scenario, noise-induced $-1$ cascades still form with low probability. This in turn means that such cascades are unlikely to form early enough to grow to an extent which would enable them to compete with early-formed $+1$ cascades. Because of this, perturbed consensus is no longer observed in the simulations.

            \paragraph{Respective influences of $\tau$ and $\zeta^+$} As noted before, for the scenario at hand, increasing either of the duration of commitments or the number of committed agents yields faster and unperturbed consensus. Consequently, one question of interest would pertain to which of these two influences is the strongest. To answer this, we note that the average network disagreement $\bar{D}$ follows a sharp monotonic transition between delayed and early consensus as the commitment duration $\tau$ increases, with a clear threshold around $\tau = 5$ for $\zeta^+ = 0.02$. Furthermore, we note that this threshold decreases as the fractions of committed agents $\zeta^+$ increase, as shown by the receding boundary between non-zero and zero disagreement values in \fgref{DDM_RR_single_Dmap}, progressively going from $\tau = 5$ to $\tau = 1$ as $\zeta^+$ increases. 

            However, this overall disagreement decrease seems to be much sharper in the direction of increasing $\tau$ than it is in the direction of increasing $\zeta^+$, suggesting that the duration of commitments plays a more significant role than the number of committed agents in facilitating an information cascade that leads to consensus. In fact, we surmise that this critical role of $\tau$ comes as a consequence of the evidence accumulation and decay mechanisms, which are typically absent in existing models of contagion. In particular, for complex contagion models \cite{YZM2021} and independent cascade models \cite{BAA2011}, the increased spread of contagion is linked to increasing the number of committed agents but not necessarily to increasing the length of commitments. In simple contagion models \cite{CPVPG2016}, the key role is played by the basic reproduction number (parameters associated with the likelihood of contagion transmission and recovery), but contagion spread is not substantially influenced by neither the number nor duration of neighbouring infections.

            As a summary of this first simulated scenario, we conclude that the introduction of a single type of information amidst the agent population overwhelmingly yields consensus. Both longer commitments and (marginally) increasing the amount of committed agents favour the fast emergence of consensus, and decrease the number of instances of perturbed consensus.

        \subsection{Opposing information sources}\label{sc:results_dual}

            \begin{figure}
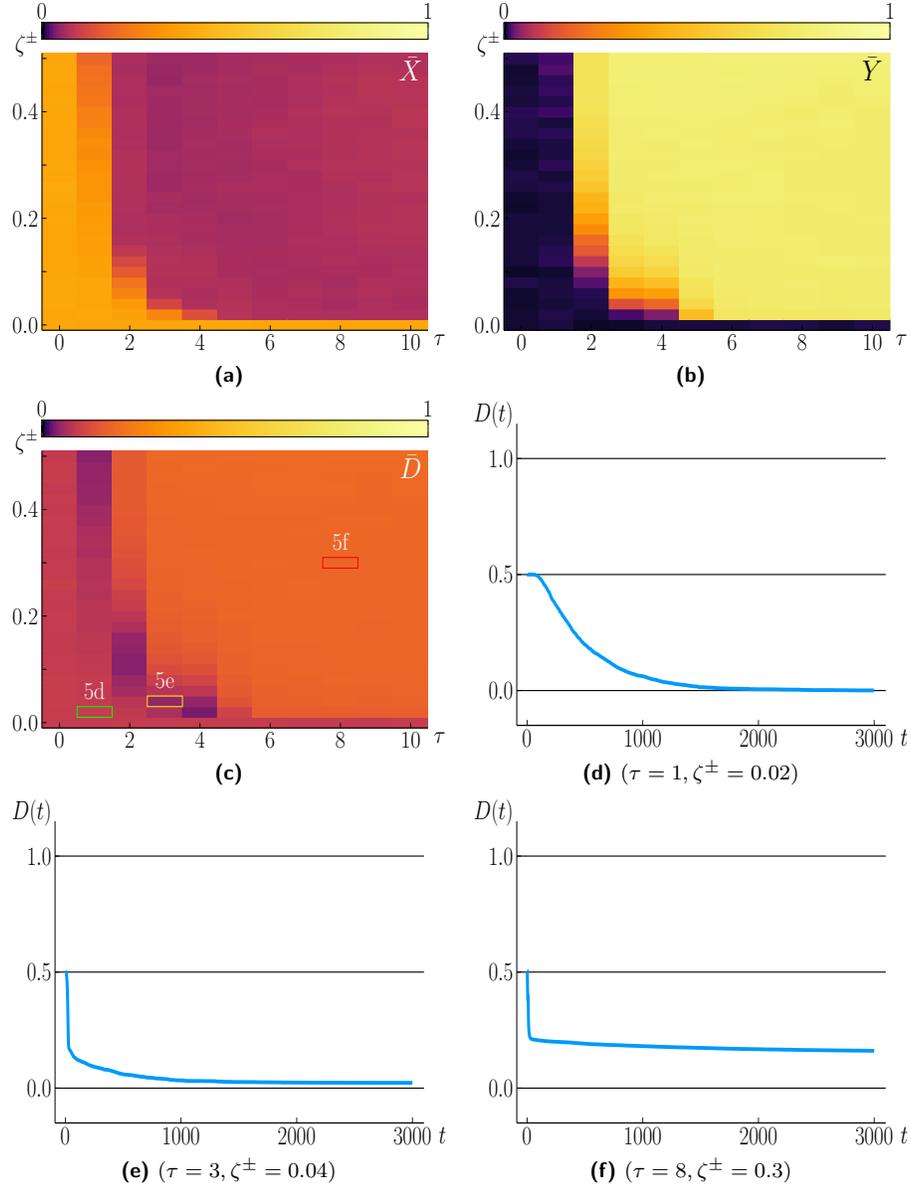

                \begin{subfigure}[t]{.5\textwidth}

                    \resizebox{\textwidth}{.8\textwidth}{\tikzsetnextfilename{DDM_RR_dual_Xmap} \input{DDM_RR_dual_Xmap.tex}}\\
                    \caption{}
                    \label{fg:DDM_RR_dual_Xmap}

                \end{subfigure}~\begin{subfigure}[t]{.5\textwidth}

                    \resizebox{\textwidth}{.8\textwidth}{\tikzsetnextfilename{DDM_RR_dual_Ymap} \input{DDM_RR_dual_Ymap.tex}}\\
                    \caption{}
                    \label{fg:DDM_RR_dual_Ymap}

                \end{subfigure}

                \begin{subfigure}[t]{.5\textwidth}

                    \resizebox{\textwidth}{.8\textwidth}{\tikzsetnextfilename{DDM_RR_dual_Dmap} \input{DDM_RR_dual_Dmap.tex}}\\
                    \caption{}
                    \label{fg:DDM_RR_dual_Dmap}

                \end{subfigure}~\begin{subfigure}[t]{.5\textwidth}

                    \resizebox{\textwidth}{.8\textwidth}{\tikzsetnextfilename{DDM_RR_dual_Dt_low} \input{DDM_RR_dual_Dt_low.tex}}\\
                    \caption{$(\tau = 1, \zeta^\pm = 0.02)$}
                    \label{fg:DDM_RR_dual_Dt_low}

                \end{subfigure}

                \begin{subfigure}[t]{.5\textwidth}

                    \resizebox{\textwidth}{.8\textwidth}{\tikzsetnextfilename{DDM_RR_dual_Dt_mid} \input{DDM_RR_dual_Dt_mid.tex}}\\
                    \caption{$(\tau = 3, \zeta^\pm = 0.04)$}
                    \label{fg:DDM_RR_dual_Dt_mid}

                \end{subfigure}~\begin{subfigure}[t]{.5\textwidth}

                    \resizebox{\textwidth}{.8\textwidth}{\tikzsetnextfilename{DDM_RR_dual_Dt_high} \input{DDM_RR_dual_Dt_high.tex}}\\
                    \caption{$(\tau = 8, \zeta^\pm = 0.3)$}
                    \label{fg:DDM_RR_dual_Dt_high}

                \end{subfigure}

                \caption{Results of the Monte Carlo simulations for a RR network in the dual-source case with equal proportions of committed agents of each type ($\zeta^- = \zeta^+ = \zeta^\pm$). In panels {\bf(a)}, {\bf(b)}, and {\bf(c)}, we report the values of the temporal entropy $\bar X$, spatial entropy $\bar Y$, and average network disagreement $\bar D$, respectively, using a colour code, for different agent commitment durations $\tau$ and fractions of committed agents $\zeta^+$. In panels {\bf(d--f)}, we represent the temporal evolution of the network disagreement $D(t)$ for three pairs of meta-parameters $(\tau,\zeta^+)$ of interest,  represented by the boxed areas in panel {\bf(c)}.}
                \label{fg:DDM_RR_dual}

            \end{figure}

            Next, as described in \scref{opposing_sources}, we study the effect of information sources of opposite types being concurrently introduced in a RR network. Following the single-source case, a natural question to ask pertains to how the introduction of a new type of agent commitment affects the emergent behaviours previously observed in \scref{results_single}. To answer this, and specifically to facilitate a comparison between the single- and dual-source cases, we find it useful to start our analysis by focusing on a sub-scenario featuring equal amounts of both types of committed agents ($\zeta^+ = \zeta^- := \zeta^\pm$). Thanks to this dimensional reduction, \fgref{DDM_RR_dual} is able to display the same metrics as \fgref{DDM_RR_single}, for the same meta-parameter region. From this, we see that \fgref{DDM_RR_single} and \fgref{DDM_RR_dual} present similar regions of interest in terms of the meta-parameters considered, allowing us to repeat our split analysis of the results into cases of weak and strong agent commitments.

            \paragraph{Weak agent commitments} For low values of $\tau$ and $\zeta^\pm$, all of our measurements bear a strong similarity to the single-source scenario. In particular, \fgref{DDM_RR_single_Dt_low} and \fgref{DDM_RR_dual_Dt_low} essentially present the same evolution of $D(t)$ for the same region of the ($\zeta^+, \tau$) plane, and the heatmaps of $(\bar{X}, \bar{Y}, \bar{D})$ do not differ significantly in their low $\tau$ and low $\zeta^+$ regions between \fgref{DDM_RR_single} and \fgref{DDM_RR_dual}. Thus, the same conclusions can be drawn for weak agent commitments of both the single- and dual-source cases: namely, that weak agent commitments yield delayed and perturbed consensus (even when two types of information are introduced in the network).

            Because of this behavioural consistency between the single- and dual-source scenarios, it is natural to make sense of this new observation by extending the rationale given for weak agent commitments in \scref{results_single}. Specifically, by conjecturing that both $-1$ and $+1$ information cascades have a low (though now equal) probability of forming at any time $t$. In this case, competition between information cascades of opposite types remains possible. In turn, the earliest cascade to form will likely yield consensus, leading to the observed instances of delayed (and sometimes perturbed) consensus.

            \paragraph{Strong agent commitments} If we extend our comparison with the single-source case to the domain of strong agent commitments, significant differences start to show. First, the zero-disagreement region of \fgref{DDM_RR_single_Dmap} has now been replaced by a region of fixed, nonzero disagreement in \fgref{DDM_RR_dual_Dmap}. Having a closer look at the evolution of $D(t)$ for this area in \fgref{DDM_RR_dual_Dt_high} reveals that the network disagreement now converges to a fixed value $D^* \approx 0.33 >> 0$ as $t$ grows large. Second, in addition of now being nonzero, spatial entropy $\bar{Y}$ now dominates temporal entropy $\bar{X}$, as shown by comparing the high $\tau$ or high $\zeta^\pm$ regions of \fgref{DDM_RR_dual_Ymap} and \fgref{DDM_RR_dual_Xmap}. As pointed out in \scref{metrics}, all of these observations suggest that polarisation is the consistent and dominant behaviour for instances of strong agent commitments. Moreover, as commitments grow stronger, this polarisation effect is made more stable over time (as shown by $D(t)$ quickly converging to a fixed value in \fgref{DDM_RR_dual_Dt_mid} and \fgref{DDM_RR_dual_Dt_high}), and involves growing numbers of disagreeing neighbours (as shown by the specific value of $D^*$ progressively increasing from a negligible value in \fgref{DDM_RR_dual_Dt_mid}, to a value of $0.33$ in \fgref{DDM_RR_dual_Dt_high}).

            Building upon a similar reasoning as before, we claim that when commitments are strong, both types of information cascades have a high probability of forming within the simulation time window. In this case, such information cascades are more likely to form within short durations of each other compared with weak agent commitments. Because of this, and because committed agents are randomly placed in the network, such cascades are also likely to form sufficiently far from each other to allow for significant concurrent growth before confrontation occurs. Just as for the single-source case, agents at the interface of such cascades will convert to match the locally dominant stance -- however, given the large and comparable sizes of cascades in this sub-scenario, this conversion process is likely to now take much longer, or stop for the rest of the simulation time window (the mathematical intuition for which is provided under \eqref{equilibrium}). In turn, agents on the boundaries of each cascade are more likely to maintain the same (non-zero) stance for a long time, leading to stable network polarisation with a fixed network disagreement $D^*$.

            \paragraph{Weak/Strong commitments boundary} The aforementioned consensus and polarisation regions are separated by a variable threshold in commitment lengths, as is visible on all heatmaps of \fgref{DDM_RR_dual}. Similarly to the single-source case, this threshold is as high as $\tau = 5$ for low amounts of committed agents, and can be observed to decrease to as low as $\tau = 1$ for high $\zeta^\pm$. Since the dominant behaviour transitions from consensus to polarisation as $\tau$ increases, and since the average network disagreement $\bar{D}$ \emph{monotonically decreased} with $\tau$ in the previous scenario, one would now expect $\bar{D}$ to \emph{monotonically increase} with $\tau$. However, a look at \fgref{DDM_RR_dual_Dmap} reveals that this is not exactly the case. In fact, as the commitment length $\tau$ increases, the average disagreement $\bar{D}$ itself seems to decrease from a low value ($\bar{D} \approx .2$)  which is typical of delayed consensus, to a \emph{local minimum} ($\bar{D} \approx 0$), before increasing to finally reach its maximum value ($\bar{D} \approx .33$) which is typical of polarisation. Interestingly, this non-monotonic transition seems to happen no matter the fraction of committed agents at play. A closer look at \fgref{DDM_RR_dual_Dt_mid} suggests for the transitory behaviour to be a perturbed kind of polarisation, where the definitive boundaries of polarised groups are not decided (slowly decreasing $D(t)$) for a significant time, before fully settling on a very low number of connected disagreeing agents ($D(t) = D^*$), hinting at a quasi-consensus situation. Interestingly, this gradual decrease of $D(t)$ is much faster than the one leading to delayed consensus for weak agent commitments -- meaning that no matter how many committed agents of both types are introduced, there exists an optimal commitment length which allows for a fast quasi-consensus.

            \paragraph{Unequal proportions of information sources} For completeness, we still quantify the impact of committed populations of unequal sizes. To this end, we turn to \fgref{DDM_RR_uneven}, which focuses on measurements of the average network disagreement for unequal proportions of committed agents and fixed commitment durations.

            \begin{figure}
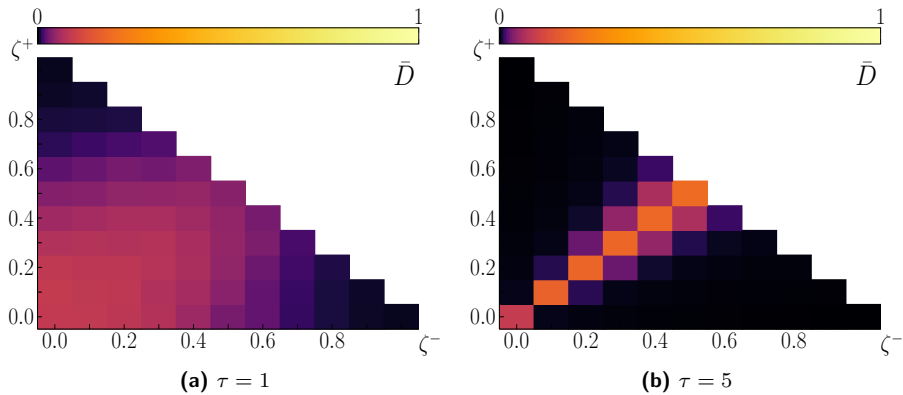


                \begin{subfigure}[t]{.5\textwidth}
    
                    \resizebox{\textwidth}{.8\textwidth}{\tikzsetnextfilename{DDM_RR_uneven_1t_Dmap} \input{DDM_RR_uneven_1t_Dmap.tex}}\\
                    \caption{$\tau = 1$}
                    \label{fg:DDM_RR_uneven_1t_Dmap}
    
                \end{subfigure}~\begin{subfigure}[t]{.5\textwidth}
    
                    \resizebox{\textwidth}{.8\textwidth}{\tikzsetnextfilename{DDM_RR_uneven_5t_Dmap} \input{DDM_RR_uneven_5t_Dmap.tex}}\\
                    \caption{$\tau = 5$}
                    \label{fg:DDM_RR_uneven_5t_Dmap}
    
                \end{subfigure}
    
                \caption{Results of the Monte Carlo simulations for a RR network in the dual-source case with unequal proportions of committed agents of each type ($\zeta^- \neq \zeta^+$). In each panel, we report the values of the average network disagreement $\bar D$ using a colour code. In panel $(\bf{a})$, the commitment duration is fixed to $\tau = 1$, while panel $(\bf{b})$ focuses on the case where $\tau = 5$.}
                \label{fg:DDM_RR_uneven}

           \end{figure}

            In this sub-scenario, the main observation is two-fold: i) the emerging behaviour seems to be symmetric with respect to the axis of equal proportions of committed agents of each type ($\zeta^- = \zeta^+$) ii) straying from this axis decreases disagreement, which we interpret as facilitating consensus. In order to explain this, we briefly note that our interpretation involving competing information cascades extends to cover this sub-scenario in a straightforward manner.

            Concluding this set of simulations, we find that the introduction of two types of information sources in the agent population still allows consensus for weak agent commitments -- but that stronger commitments instead introduce network polarisation. Moreover, the robustness of this polarising effect increases with longer and longer commitments, which serves as a good baseline observation to describe the quantitative differences observed between Random-Regular networks and other topologies.

        \subsection{Impact of the network topology}\label{sc:results_other}

            In this section, we extend the results obtained for RR networks to other network topologies, to shed light on the impact of the network structure on the emergence of the different behaviours of interest described above. Following the methodology described in \scref{other_topologies}, we first consider the dual-source scenario on RR networks with larger agent populations. Then, we consider different network topologies, namely the BA, WS, and SBM networks described in \scref{other_topologies}.

            \subsubsection{Larger RR networks}

                \begin{figure*}
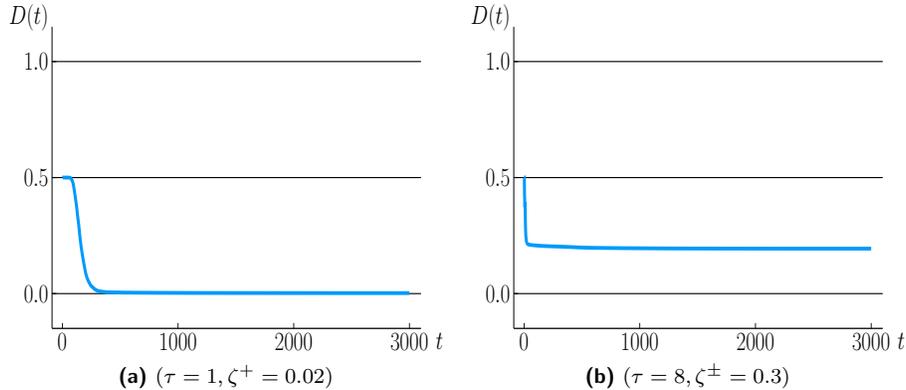


                    \begin{subfigure}[t]{.5\textwidth}

                        \resizebox{\textwidth}{.8\textwidth}{\tikzsetnextfilename{DDM_RRL_single_Dt_low} \input{DDM_RRL_single_Dt_low.tex}}\\
                        \caption{$(\tau = 1, \zeta^+ = 0.02)$}
                        \label{fg:DDM_RRL_single_Dt_low}
                    
                    \end{subfigure}~\begin{subfigure}[t]{.5\textwidth}
    
                        \resizebox{\textwidth}{.8\textwidth}{\tikzsetnextfilename{DDM_RRL_dual_Dt_high} \input{DDM_RRL_dual_Dt_high.tex}}\\
                        \caption{$(\tau = 8, \zeta^\pm = 0.3)$}
                        \label{fg:DDM_RRL_dual_Dt_high}
    
                    \end{subfigure}

                    \caption{Temporal evolution of the network disagreement $D(t)$ for a large ($N = 1000$) RR network computed via Monte Carlo simulations. Panel {\bf(a)} represents a weak-commitment region in the single-source case, while panel {\bf(b)} represents a strong-commitment region in the dual-source case with equal proportions of committed agents of both types.}
                    \label{fg:DDM_RRL}

                \end{figure*}

                For the sake of brevity, and because results do not differ significantly for this sub-scenario, \fgref{DDM_RRL} only shows network disagreement $D(t)$ for two specific pairs of meta-parameters. Specifically, \fgref{DDM_RRL_single_Dt_low} displays the same weak-commitment region of the \emph{single-source} information scenario as \fgref{DDM_RR_single_Dt_low}, while \fgref{DDM_RRL_dual_Dt_high} focuses on the strong-commitment region of the \emph{dual-source} information scenario shown in \fgref{DDM_RR_dual_Dt_high}.

                In both cases, the evolution of $D(t)$ is found to be similar as that observed for smaller RR networks, up to minor quantitative differences. For weak agents commitments (\fgref{DDM_RRL_single_Dt_low}), delayed consensus can still be inferred to occur, but the mean time needed to achieve $D(t) \to 0$ is much shorter than in smaller networks ($t \approx 300$ vs $t \approx 1700)$. 
                For strong agent commitments of opposite types (\fgref{DDM_RR_dual_Dt_high}), polarisation is also still observed, however $D(t)$ seems to converge to a higher final value ($D^* \approx 0.19$ vs $D^* \approx 0.16$). 

                In order to explain the former observation, note that for the same value of $\zeta^+$, there are ten times more committed agents for networks with $N = 1000$ than for networks with $N = 100$. Consequently, and because of our intuition on how information cascades are formed, it would make sense for the probability of an information cascade forming at any time to also follow a proportional increase in larger networks, which would explain the shorter times to consensus observed.
                Moving on to the latter observation, recalling that $D^*$ measures a proportion of how many pairs of disagreeing agents are connected by an edge also makes for an intuitive explanation of the higher values of $D^*$ observed in fully polarised larger networks.

                Thus, we conclude that results obtained for both the single-source and dual-source scenarios generalise to larger networks in a straightforward fashion, with expected quantitative differences in the main metrics considered.

            \subsubsection{Different network topologies}

                \begin{figure*}
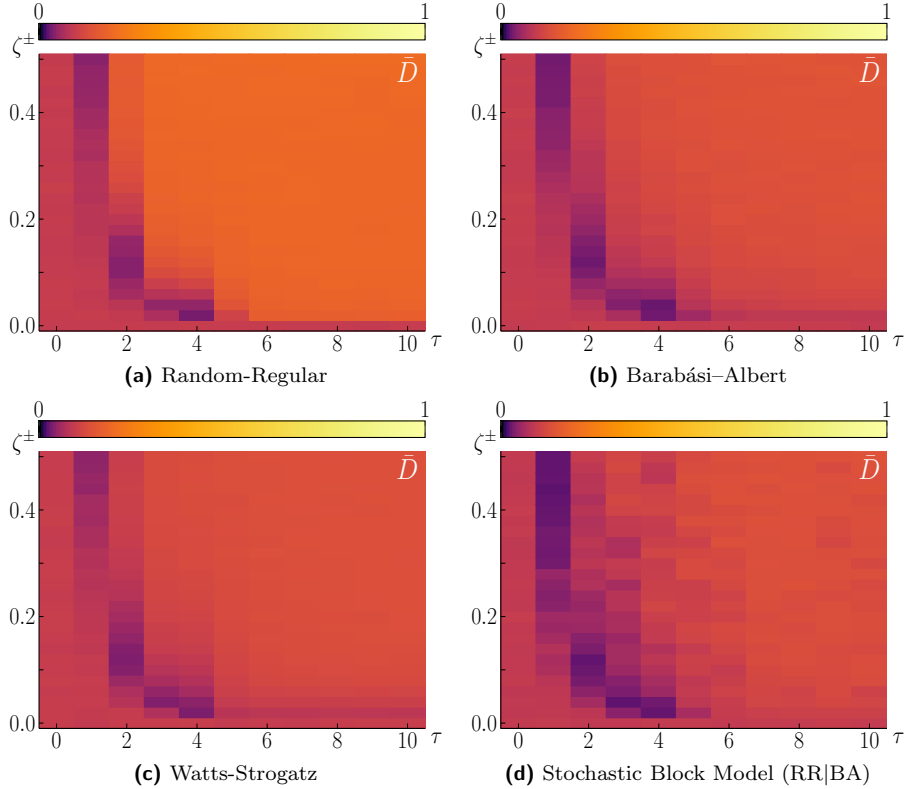


                    \begin{subfigure}[t]{.5\textwidth}

                        \resizebox{\textwidth}{.8\textwidth}{\tikzsetnextfilename{DDM_RR_dual_Dmap_bis} \input{DDM_RR_dual_Dmap_bis.tex}}\\
                        \caption{\centering Random-Regular}
                        \label{fg:DDM_RR_dual_Dmap_bis}

                    \end{subfigure}~\begin{subfigure}[t]{.5\textwidth}

                        \resizebox{\textwidth}{.8\textwidth}{\tikzsetnextfilename{DDM_BA_dual_Dmap} \input{DDM_BA_dual_Dmap.tex}}\\
                        \caption{\centering Barabási–Albert}
                        \label{fg:DDM_BA_dual_Dmap}

                    \end{subfigure}

                    \begin{subfigure}[t]{.5\textwidth}

                        \resizebox{\textwidth}{.8\textwidth}{\tikzsetnextfilename{DDM_WS_dual_Dmap} \input{DDM_WS_dual_Dmap.tex}}\\
                        \caption{\centering Watts-Strogatz}
                        \label{fg:DDM_WS_dual_Dmap}

                    \end{subfigure}~\begin{subfigure}[t]{.5\textwidth}

                        \resizebox{\textwidth}{.8\textwidth}{\tikzsetnextfilename{DDM_RRBA_dual_Dmap} \input{DDM_RRBA_dual_Dmap.tex}}\\
                        \caption{\centering Stochastic Block Model (RR$|$BA)}
                        \label{fg:DDM_RRBA_dual_Dmap}

                    \end{subfigure}

                    \caption{Average network disagreement $\bar D$ for RR, BA, WS, and SBM networks in the dual-source case with equal proportions of committed agents of each type ($\zeta^- = \zeta^+ = \zeta^\pm$) computed via Monte Carlo simulations.}
                    \label{fg:DDM_XX_dual}

                \end{figure*}

                Finally, we reproduce our results while considering agents who interact on diverse network topologies (namely, BA and WS networks and SBMs). Specifically, we focus on the case of competing information sources with equal fractions of committed agents of each type ($\zeta^- = \zeta^+ = \zeta^\pm$) (see \scref{opposing_sources}). The results obtained for all of the network topologies listed in \scref{other_topologies} are summarised in \fgref{DDM_XX_dual}, which focuses on the average network disagreement $\bar{D}$ as the main metric being compared across network topologies. For comparison purposes, the heatmap of $\bar{D}$ for small ($N = 100$) RR networks is reported once again in \fgref{DDM_RR_dual_Dmap_bis}.

                First, we notice that the simulation behaviours pointed out in \scref{results_dual} were qualitatively replicated for all networks considered, and that they follow similar transitions for similar meta-parameter values. In particular, the non-monotonicity of average network disagreement in both increasing directions of the $(\tau, \zeta^\pm)$ plane is preserved, and so is the higher influence of commitment lengths ($\tau$) over commitment numbers ($\zeta^\pm$).
                Besides these similarities, and while all networks seem to agree on a baseline $\bar{D}$ value for weak commitments, noticeable quantitative differences can be observed for strong commitments across the different network types. In particular, RR networks seem to yield higher values of average network disagreement as compared to other networks, while SBMs exhibit mixed values.

                In fact, both of these differences can be readily explained by considering the defining characteristics of each network topology at hand. Because of the uniform degree distribution of RR networks, when polarisation occurs, the boundary between polarised clusters of agents is likely to always involve the same (high, perhaps maximal) number of agents, thus yielding a high $D^*$ value. By contrast, this phenomenon is unlikely to occur in other topologies, where the degree distribution is not as uniform and information cascades are not guaranteed to split the network across a path where a lot of agents lie. In fact, considering a situation where the interface between the two polarised agent clusters comprises only one or two agents yields a rationale for the simulation behaviours observed in SBMs. Indeed, because the link probability $\lambda$ between the two sub-communities of such networks is negligible, in the extreme event of a $-1$ and a $+1$ cascade each forming in one of the two sub-communities, the network would likely end up polarised along the (few) links between said sub-communities, leading to an unusually low $D^*$ value. Thus, for SBMs, the specific observed value of $D^*$ in any simulation instance highly depends on where information cascades first form in the network.

                As a conclusion, we report that the main results presented throughout \scref{results_single} and \scref{results_dual} seem to generalise quite well to a diverse range of network topologies, despite minor quantitative discrepancies which can be explained away by differences in individual degree distributions. In particular, our key observation of the length of commitments having an higher influence than the number of committed agents on consensus/polarisation still holds, which further consolidates the hypothesis of this feature being a distinguishing consequence of introducing evidence accumulation in our model dynamics.

    \section{Conclusions}\label{sc:discussion}

        In this paper, we have introduced a model of competing information spread, that focused on evidence accumulation, using a social drift-diffusion model. Furthermore, we have developed information-theoretical and graph-based metrics to characterise emergent behaviours of the system, notably polarisation and consensus. Computing these metrics by means of Monte Carlo simulations, we have shown that the emergence of consensus or polarisation is conditional on the type, number, and persistence of information sources being introduced among the agent population. However, contrary to current models of information spread, we have found that out of these three influences, information source persistence is the stronger factor in deciding which behaviour emerges. We suspect that this is due to the evidence accumulation mechanism in our model, which adds a strong dependence of the dynamics on past agent interactions, absent from most current models of information spread.

        As leads for future works, we propose natural generalisations of our model; namely, more realistic (but more complex) versions of the model could include heterogeneous values of the parameters across the population, as these quantities would typically change on an individual basis in real-world settings. Moreover, we point out a few directions for theoretical analysis, such as to characterise the steady-state distribution of agents' states for simple cases, such as RR networks with known locations of committed agents, and to investigate the exact (seemingly exponential) nature of the $\zeta/\tau$ dependence. Finally, a rigorous real-world validation of the proposed mathematical model against real-world data should be performed.

    \bibliographystyle{elsarticle-num}
    \bibliography{P1.bib}

    \pagebreak
    \appendix

    \section{Analytical insights}\label{sc:analysis}

        To guide our exploration of the emergent behaviours of the model, we derive some mathematical insight into the dynamics, useful in restricting the model's parameter space to a sub-space of interest for our simulations experiments, as well as in explaining how information cascades behave in our model.

        \paragraph{Confidence Equilibrium} First, consider an agent~$i$ whose neighbours $j \in \mathcal{N}_i$ have constant stances for all times $t \in \NN$. Let $N^+$ denote the number of such $+1$ stances, and $N^-$ denote the number of such $-1$ stances, so that the net stance of $\mathcal{N}_i$ can be expressed as $\sum_{j\in\mathcal{N}_i} s_j(t) = N^+ - N^-$. Because of \eqref{confidence_update}, if this net stance is nonzero, we expect the confidence of agent $i$ to keep shifting by the value of said net stance (up to noise perturbations). Consequently, the \emph{expected confidence} of agent $i$ has a definite limiting behaviour, which we now derive. Rearranging \eqref{confidence_update}, then taking the expected value of both sides yields the following recurrence relation:
        \begin{equation*}
            \EE[c_i(t+1)] = (1 - \delta)\EE[c_i(t)] + N^+ - N^- + \delta\beta_i,
        \end{equation*}
        which yields the following limit for $0 < \delta \leq 1$:
        \begin{equation}\label{eq:equilibrium}
            \lim_{t \to \infty}{\EE[c_i(t)]} = (N^+ - N^-)\frac{1}{\delta} + \beta_i.
        \end{equation}
        Thus, \eqref{equilibrium} implies that the long-term expected confidence of agent $i$ is a linear function of the net stance of their neighbours. As a consequence of this, for reasonable values of model parameters (e.g. satisfying $\rho < \frac{1}{\delta} + |\beta_i|$), agent $i$'s long-term stance is essentially decided by a majority rule. Conversely, this explains how a single agent can initiate information cascades in the network; namely by having a constant nonzero stance for long enough to start a confidence feedback loop between neighbouring agents. Furthermore, \eqref{equilibrium} also provides insight into what can possibly happen when two information cascades of opposite stances meet in a network. As implied above, agents at the interface of such two cascades will, given enough time, convert to the stance that locally prevails in their neighbourhood. Iterating this process of successive agent conversion at the interface of the two cascades ends in two possible ways: i) The process continues until all agents are of the same stance, yielding consensus or ii) the process stops after a certain number of ``inconvertible'' neighbourhoods with zero net stance are created, yielding sustained polarisation.

        Note that the duration of this conversion process highly depends on how large agent confidences have been allowed to grow prior to cascades meeting, which is intuitively analogous to how long agents have been part of said information cascades. Quantifying this further, \eqref{equilibrium} highlights that the maximum expected agent confidence scales with $\delta^{-1}$. Consequently, for reasonable values of $\delta$, any agent who has spent a lot of time partaking in an information cascade only has a negligible probability to ever revert to not sharing (or sharing opposite information) -- a phenomenon which could be related to echo chambers, often observed in real-world social networks \cite{CDFMG2021}

        \paragraph{Stance continuity} The previous result implies that information cascades are formed whenever agent stances are consistent over a minimal time window and set of agents. However, due to the stochasticity of our model, it is theoretically possible for any agent's stance to change solely due to evidence noise. In order to quantify how likely this phenomenon is based on model parameters, we now derive a bound on the probability that an agent's stance takes opposite values in two consecutive time steps. To this end, we partition such an event as $A \cup B$ with:
        \begin{equation*}
            \begin{cases}
                s_i(t+1) = +1 \st s_i(t) = -1 & (A) \\ 
                s_i(t+1) = -1 \st s_i(t) = +1 & (B). \\
            \end{cases}
        \end{equation*}

        Starting with the first sub-event, notice that $\PP{A}$ is maximised by agent $i$ having an a priori confidence $c_i(t)$ of exactly $-\rho$, and all $|\mathcal{N}_i|$ of its neighbours having an a priori stance of $+1$. Substituting these values in \eqref{confidence_update} and \eqref{stance_update}, and noticing that $\delta = 1$ maximises the resulting bound yields
        \begin{equation*}
            \PP{A} \enskip\leq\enskip \PP{e_i(t) > (2 - \delta)\rho - \delta|\beta_i| - |\mathcal{N}_i|} \enskip\leq\enskip \Phi\left(\frac{|\mathcal{N}_i| + |\beta_i| - \rho}{\sigma}\right),
        \end{equation*}
        where $\Phi(x)$ is the cumulative distribution function of a standard normal distribution~\cite{Ross2023}. Applying a similar reasoning to event $B$, we find that the same bound also applies to $\PP{B}$, allowing us to achieve the following bound on the total probability of our original event:
        \begin{equation}\label{eq:switchpr_bound}
            \PP{s_i(t+1)s_i(t) = -1} \enskip\leq\enskip 2\Phi\left(\frac{|\mathcal{N}_i| + |\beta_i| - \rho}{\sigma}\right).
        \end{equation}

        Thanks to this result, the probability that a non-committed agent switches the stance of the information they share in a single time step can be made extremely small by carefully choosing model parameters. In particular, for unbiased agents ($\beta_i = 0$) in a network of average degree $K$, setting $\rho > K$ is enough to ensure that the average agent is unlikely to oscillate between opposite stances, in turn ensuring evidence accumulation remains central to the formation of information cascades, without being overshadowed by evidence noise.

\end{document}